\documentclass[%
reprint,
superscriptaddress,
amsmath,amssymb,
aps,
pre,
]{revtex4-1}

\usepackage{graphicx}
\usepackage{float}
\usepackage[caption=false]{subfig}
\usepackage{dcolumn}
\usepackage{bm}

\begin{document}
\preprint{APS/123-QED}
\title{The effect of RNA stiffness on the self-assembly of virus particles}

\author{Siyu Li}
\affiliation{Department of Physics and Astronomy,
University of California, Riverside, California 92521, USA}
\author{Gonca Erdemci-Tandogan}
\affiliation{Department of Physics, Syracuse University, Syracuse, NY 13244, USA}
\author{Paul van der Schoot}
\affiliation{Group Theory of Polymers and Soft Matter, Eindhoven University of Technology, P.O. Box 513, 5600 MB Eindhoven,
The Netherlands}
 \affiliation{ Institute for Theoretical Physics,
   Utrecht University,
   Princetonplein 5, 3584 CC  Utrecht, The Netherlands}
\author{Roya Zandi}
\affiliation{Department of Physics and Astronomy,
   University of California, Riverside, California 92521, USA}





\date{\today}

\begin{abstract}
Under many {\it in vitro} conditions, some small viruses spontaneously encapsidate a single stranded (ss) RNA into a protein shell called the capsid. While viral RNAs are found to be compact and highly branched because of long distance base-pairing between nucleotides, recent experiments reveal that in a head-to-head competition between a ssRNA with no secondary or higher order structure and a viral RNA, the capsid proteins preferentially encapsulate the linear polymer! In this paper, we study the impact of genome stiffness on the encapsidation free energy of the complex of RNA and capsid proteins. We show that an increase in effective chain stiffness because of base-pairing could be the reason why under certain conditions linear chains have an advantage over branched chains when it comes to encapsidation efficiency. While branching makes the genome more compact, RNA base-pairing increases the effective Kuhn length of the RNA molecule, which could result in an increase of the free energy of RNA confinement, that is, the work required to encapsidate RNA, and thus less efficient packaging.
\end{abstract}

\maketitle


\section{introduction}

Ribonucleic acid (RNA) is one of the molecules of life, which plays a central role in the cell as information carriers, enzymes, gene regulators, et cetera. It is made out of four elementary building nucleotides, being A(denine), G(uanine), C(ytosine) and U(racil) \cite{Higgs2000}. As shown by Crick and Watson, purines (A,G) pair with complementary pyrimidines (C,U), leading primarily to the pairs CG and AU. There exist also so-called wobble pairs of GU. Single stranded RNA is quite flexible with a Kuhn length of, depending on the ionic strength of the solution, one or two $nm$ \cite{Chen2011}, and can form double helical stems (A helices) with a Kuhn length of about 140 $nm$ \cite{Kebbekus1995,Abels2005}. So, double stranded RNA is stiffer than double stranded DNA, which has a Kuhn length of 100 $nm$, noting that the Kuhn length is twice the persistence length of a so-valled wormlike chain.

The pairing of bases over long distances along the backbone gives rise to the secondary or folded structure of RNA. Pairing of bases can be represented by so-called arch diagrams. Nested arches represent helices, while crossings give rise to the so-called pseudoknots \cite{Nussinov}. The nested pairings can be described quantitatively by recursion relations \cite{McCaskill1990,Zuker,Vienna}, which exactly sum all possible pairings without pseudoknots. From a geometrical point of view, the generated structures can be viewed as branched polymers. The size of an ideal, Gaussian linear polymer scales as the number of ``segments'' to the power $\nu=1/2$, while ideal branched ones have a scaling exponent $\nu=1/4$ \cite{Schwab2009}. Note that there is no excluded volume interaction between monomers of an ideal chain. For self-avoiding chains the scaling exponents are $\nu=3/5$ and $\nu=1/2$ for the linear and branched polymers, respectively \cite{Grosberg97,Schwab2009}. However, because of its \textit{tertiary} structures that include pseudoknots, RNAs are significantly more compact than branched polymers. Indeed, several numerical studies and surveys have found the exponent $\nu=1/3$ to be small for RNA, reflecting this more compact structure\cite{Fang2011,Ben-Shaul2015}.

Many small viruses encapsidate a single stranded RNA into a protein shell called the capsid. Under appropriate physico-chemical conditions of acidity and ionic strength, this process is spontaneous and the virus can readily assemble {\it in vitro} from solutions containing protein subunits and RNA \cite{Cornelissen2007,Ren2006,Bogdan,Anze2,Zlotnick,Sun2007,Nature2016}. Note that in the absence of genome, capsids do not form at physiological pH and salt concentrations. Many spherical viruses adopt structures with icosahedral symmetry \cite{Fejer:10,Rapaport:04a}, which imposes a constraint on the number of subunits in capsids. The structural index $T$, introduced by Casper and Klug, defines the number of protein subunits in viral shells, which is 60 times the $T$ number. Note that $T=1,3,4,7,\ldots$ can assume only certain ``magic'' integer numbers \cite{Wagner2015956,Luque:2010a,Chen:2007b,Stefan}.

Quite interestingly, virus protein subunits are able to co-assemble with a wide variety of negatively charged cargos, including non-cognate RNAs of different length and sequence, synthetic polyanions, and negatively charged nanoparticles \cite{Sun2007,Kusters2015,Zandi2016}. It is now widely accepted that electrostatic interactions between the positive charges on the coat protein tails and negative charges on the cargo is the main driving force for the spontaneous assembly of simple viruses in solution \cite{Cornelissen2007,Ren2006,Bogdan,Anze2,Zlotnick,Hsiang-Ku,Venky2016}. Still, several recent self-assembly experimental studies reveal the importance of non-electrostatic interactions, associated with specific structures of the genome, for the selection of one RNA over another by the capsid proteins\cite{Patel2014}.

The self-assembly studies of Comas-Garcia {\it et al.}~\cite{Comas} reveal in particular the importance of RNA topology. They carried out a number of experiments in which a solution of the capsid proteins of cowpea chlorotic mottle virus(CCMV) were mixed with equal amount of RNA1 of Brome Mosaic virus (BMV) and RNA1 of Cowpea Chlorotic Mottle Virus (CCMV). In this head-to-head competition, the amount of coat protein (CP) of CCMV was selected such that it could only encapsidate one of the genomes. Quite unexpectedly, the RNA1 of CCMV (the cognate RNA) lost to RNA1 of BMV, {\it i.e}, only RNA1 of BMV was encapsidated by CCMV CPs. These experiments emphasize the impact of RNA structure on the assembly of viral shells, as RNA1 of BMV has a more compact structure than that of CCMV \cite{Gonca2014}.

Following these experiments a number of simulation studies, using quenched (fixed) branched polymers as a model for RNA, have shown that the optimal length of encapsidated RNA increases when accounting for its secondary structure \cite{elife,Ben-Shaul2015}. Mean-field calculations using annealed (equilibrium) branched polymers as model RNAs have also shown that the length of encapsidated polymer increases as the propensity to form larger numbers of branched points increases \cite{Gonca2014,Gonca2016,Li2017}. More importantly, these calculations show that a higher level of branching considerably increases the depth of the free-energy gain associated with the encapsulation of RNA by a positively charged shell. This implies that the efficiency of genome packaging goes up with increasing the level of branching, so with increasing compact secondary structure of the genome.

\begin{figure} 
   \centering
   \subfloat[]{\includegraphics[width=0.2\textwidth]{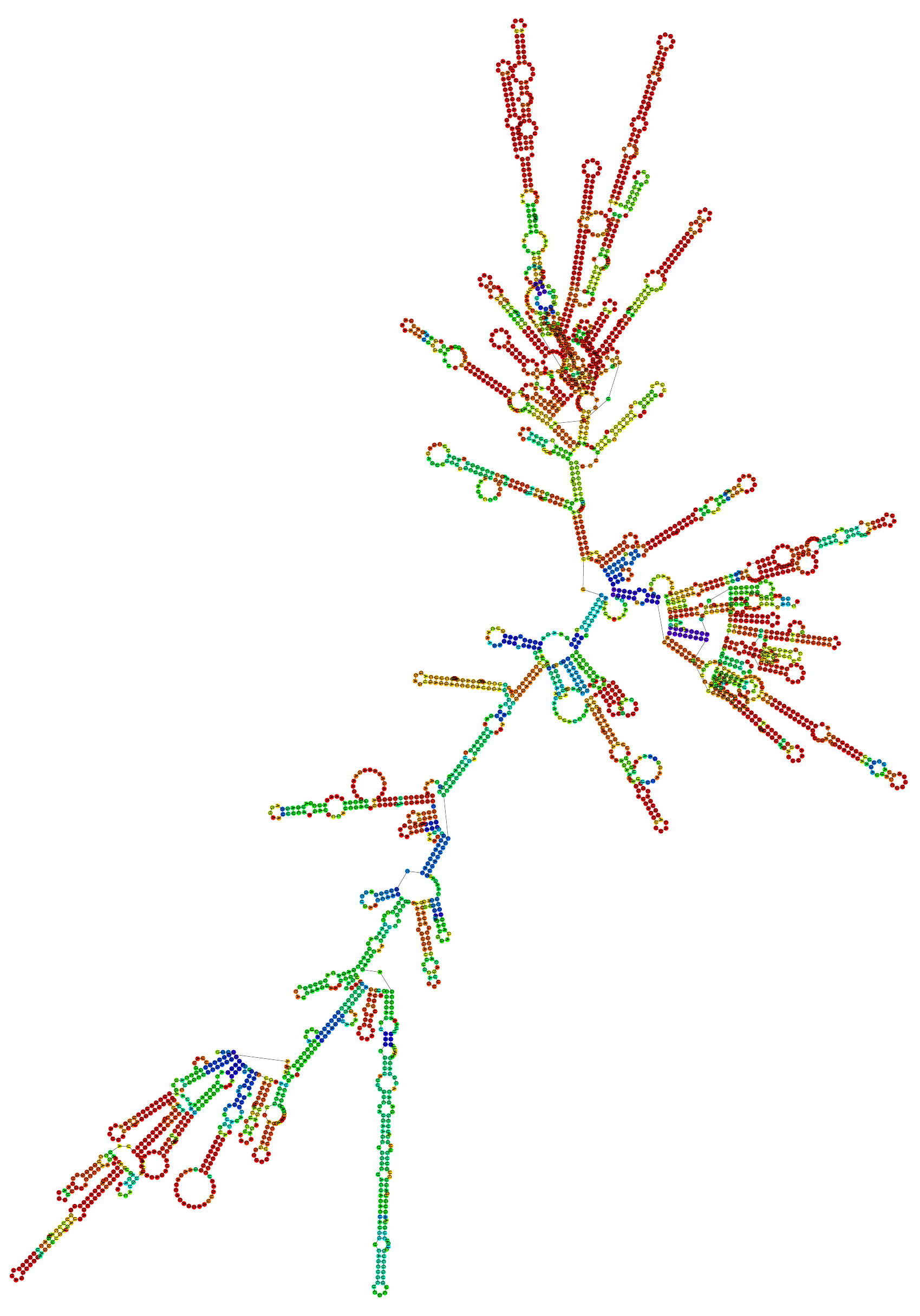}}\hspace{0.5cm}
   \subfloat[]{\includegraphics[width=0.2\textwidth]{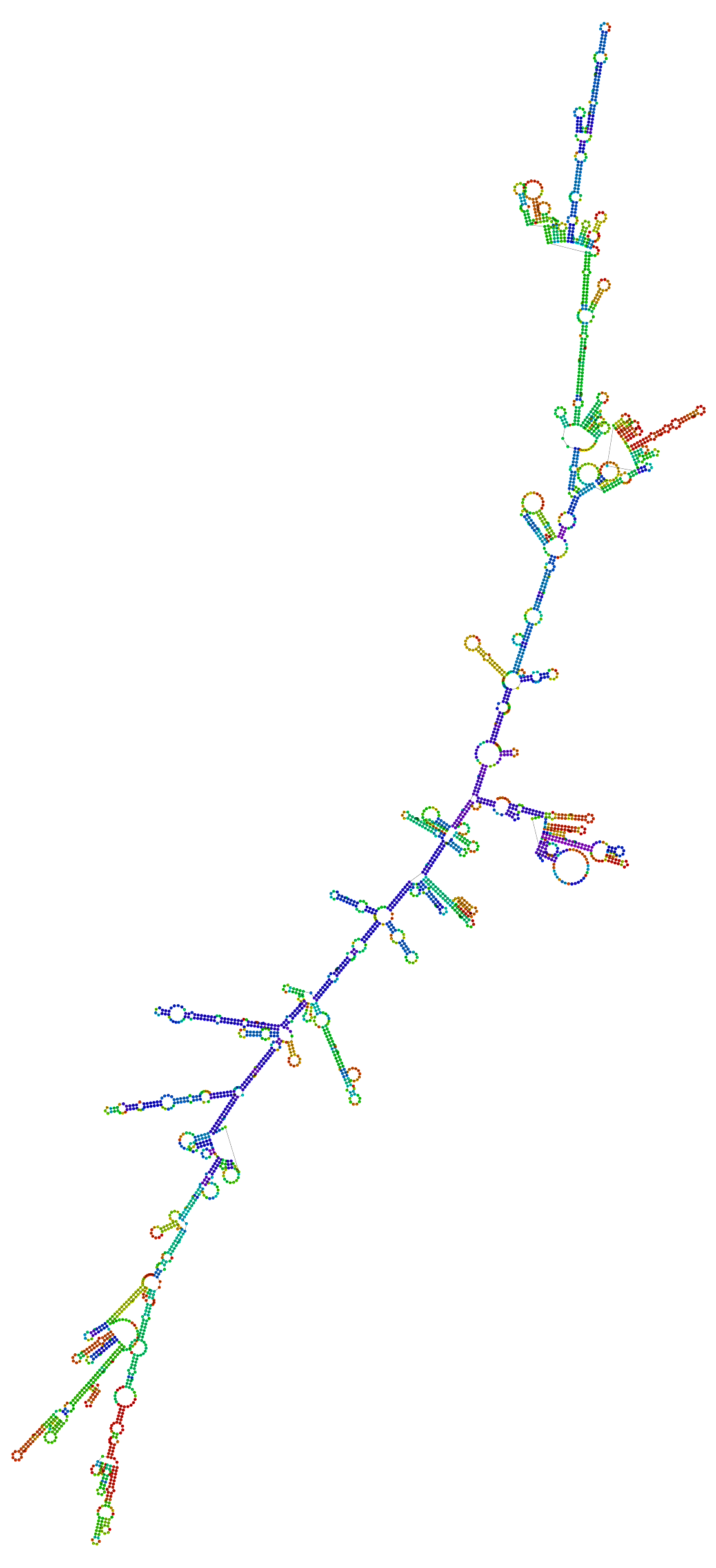}}
   \caption{\footnotesize (a) The secondary structure of the CCMV RNA1 and (b) a random RNA with the same number of nucleotides. The structures are obtained using the the Vienna RNA package \cite{Vienna}.}
   \label{structure}
\end{figure}

In fact, it was shown in Refs.~\cite{Yoffe2008,Bruinsma2016} that while RNA molecules of the same nucleotide length and composition might have similar amounts of base pairing, non-viral RNAs have significantly less compact structures than viral ones. The compactness of viral RNAs has been associated with the presence of a larger fraction of higher-order junctions or branch points in their secondary structure \cite{Yoffe2008,Li-tai,Gopal2014}.  Figure \ref{structure}(a) and (b) illustrate the secondary structures of CCMV RNA and those of a randomly sequenced RNA with the same length. The structures are obtained through the Vienna RNA software package \cite{Vienna}.  As shown in the figure, CCMV RNA has considerably larger number of branched points than non-viral RNA of the same length.

Above-mentioned theoretical and experimental studies indicate that in a head-to-head competition between two different types of RNAs, the RNA with a larger number of branching junctions or branch points should have a competitive edge over others \cite{Gonca2014,Gonca2016,Li2017}. A naive physical explanation is that branching causes RNA molecules to become more compact than structureless linear polymers of similar chain length, which are then easier to accommodate in the limited space provided by the cavity of a capsid. According to these theories and simulations, a linear chain should definitely ``loose'' to a branched one of the same number of monomers when competing head-to-head for a limited number of capsid proteins.

To probe the effect of RNA structure and test the above theories on the self-assembly of virions more systematically, Beren \textit{et al.} \cite{Beren2017} recently performed a set of {\it in vitro} packaging experiments with polyU, an RNA molecule that has no folded secondary structure. They examined whether RNA topology, i.e., the secondary structure or level of branching, allows the viral RNA to be exclusively packaged by its cognate capsid proteins. More specifically, they studied the competition between CCMV viral RNA with polyU of equal number of nucleotides for virus capsid proteins. They find that CCMV CPs are capable of packaging polyU RNAs and, quite interestingly, polyU outcompetes the native CCMV RNA in a head-to-head competition for the capsid proteins. These findings are in sharp contrast with the previous experimental, theoretical, simulation and scaling studies noted above, which suggest that the branching and compactness of RNA must lead to a more efficient capsid assembly. That being said, the scaling theory of Ref.~\cite{Paul:13a} already hints at the subtle interplay between Kuhn length, solvent quality and linear charge density dictating the free energy gain of encapsulation.

To explain these intriguing experimental findings, we employ a mean-field density functional theory and study the impact of RNA branching, while allowing for differences in Kuhn length. We further consider that double helical sequences have a larger linear charge density than non-hybridized sequences along the chain. In all previous theoretical and simulation studies related to the impact of RNA topology on virus assembly, the focus has been on the importance of the degree of branching, ignoring the impact of base-pairing on the RNA Kuhn length and linear charge density.

As noted above, the Kuhn length of single stranded RNA under physiological conditions of monovalent salt is between one and two $nm$ depending on the ionic strength \cite{Chen2011}, while that of a double stranded RNA is about 140 $nm$ \cite{Kebbekus1995,Abels2005}. The average duplex length of viral RNA is about six nucleotide pairs \cite{Fang2011}, which corresponds to about five $nm$. This value is much smaller than the persistence length of double stranded RNA\cite{Yoffe2008}, suggesting that viral RNA can be modeled as a flexible polymer with an average Kuhn length of about six paired nucleotides. There are of course also loop sequences that in our model act as end, hinge and branching points, but how this translates into an effective Kuhn length for the entire branched chain representation of the RNA is unclear. Plausibly, the effective Kuhn length of the internally hybridized chain should be larger than that of the equivalent unstructured non-hybridized chain. Furthermore, another major difference between the linear and branched (base-paired) ssRNA structures seems to be the linear charge density, which doubles for the latter on account of base pairing (hybredization).

In this paper, we vary the degree of branching as well as the effective Kuhn length and linear charge density of a model RNA, and study their impact on the optimal length of encapsulated genome by capsid proteins.  We find that as we increase the chain stiffness or Kuhn length the free energy of encapsulation of RNA becomes less negative than that of a linear chain, at least under certain conditions. Hence, a larger Kuhn length, associated with base-pairing, might decrease the efficiency of packaging of RNA compared to a linear polymer.  In contrast, our results indicate that increasing the linear charge density improves the efficiency of packaging of both linear and branched polymers.  Thus base-pairing has two competing effects: it makes the chain stiffer, which increases the work required to encapsidate the chain, but at the same time it increases the linear charge density that lowers the encapsidation free energy and augment the packaging efficiency. These results are consistent with the experiments of Beren {\it et al.} \cite{Beren2017}, in which the linear RNA, PolyU, outcompetes the cognate RNA of CCMV when they are both in solution with a limited amount of capsid proteins of CCMV, that is, sufficient to encapsidate either PolyU or CCMV RNA but not both.

The remainder of this paper is organized as follows. In the next section, we introduce the model and present the equations that we will employ later. In Section III, we present our results and discuss the impact of the Kuhn length on the capsid stability and optimal length of encapsidated genome in Section IV. Finally, in Section V, we present our conclusion and summarize our findings.
\section{Model}
To obtain the free energy associated with a genome trapped inside a spherical capsid, we consider RNA as a generic flexible
branched polyelectrolyte that interacts with positive charges residing on the inner surface of the capsid.  We focus on the case of annealed branched polymers as the degree of branching of RNAs, a statistical quantity, can be modified by its interaction with the positive charges on the capid proteins \cite{McPherson}.
Within mean-field theory, the free energy of a
negatively charged chain in a salt solution confined inside a positively charged spherical shell can be written as \cite{Borukhov,Gonca2014,Gonca2016,adsorption2015,Erdemci2016,Venky2016,Li2017}
\begin{multline} \label{free_energy}
  \beta F = \!\!\int\!\! {\mathrm{d}^3}{{\mathbf{r}}}\Big[
    \tfrac{\ell^2}{6} |{\nabla\Psi({\bf{r}})}|^2
     +\frac{1}{2}\upsilon \Psi^4 ({\mathbf{r}})+W\big[\Psi({\bf{r}}) \big]\\
    -\tfrac{ 1}{8 \pi \lambda_B} |{\nabla \beta e \Phi({\bf{r}})}|^2
    -2\mu\cosh\big[\beta e \Phi({\bf{r}})\big]
    + \beta \tau \Phi({\bf{r}})\Psi^2({\bf{r}})
    \Big]\\
  + \int\!\! {\mathrm{d}^2}r \Big[ \beta \sigma \, \Phi({\bf{r}}) \Big].
\end{multline}
with $\beta$ the inverse of temperature in units of energy, $v$ the effective excluded volume per monomer, $\lambda_B=e^2 \beta/4 \pi \epsilon$ the Bjerrum length, $e$ the elementary charge, $\mu$ the number density of monovalent salt ions, and $\tau$ the charge of the statistical Kuhn segment of the chain. The dielectric permittivity of the medium $\epsilon$ is assumed to be constant \cite{Janssen2014}.  The quantity $\ell$, the Kuhn length of the polymer, is defined as an effective stiffness averaged over the entire sequence along the genome. Further, the fields $\Psi(r)$ and $\Phi(r)$ describe the square root of the monomer density field and the electrostatic potential, respectively, and the term $W[\Psi]$ corresponds to the free energy density of an annealed branched polymer as described in Eq.~\ref{W_branched} below.

As discussed in the Introduction, the secondary structure of the RNA molecules contain considerable numbers of junctions of single-stranded loops from which three or more duplexes exit. This makes RNA act effectively as a flexible branched polymer in solution. While the Kuhn length for a single stranded, non self-hybridized ssRNA is a few nanometers and that for a double stranded RNA is about 140 nanometers, the Kuhn length of viral RNA is not well determined, as we discussed above. In the absence of exact measurements, we employ an average or effective value for $\ell$, which presumably will be larger if the number of consecutive base pairs (duplexes) between single stranded segments or stem loops along the RNA is larger. Further, we consider the limit of long chains consisting of a very large number of segments $N \rightarrow \infty$ for our confined chains, where $N$ denotes the number of segments. In this formal limit, we employ the ground-state dominance approximation implicit in Eq. (1), as it has proven to be accurate provided $N \gg 1$, i.e., for very long chains \cite{deGennes1979}. We specify below the connection between the number of segments and the number of nucleotides that make up the RNA, differentiating between self-hybridized and non self-hybridized RNAs.

The first term in Eq.~\eqref{free_energy} is the entropic cost of deviation from a uniform chain density and the second term describes the influence of excluded volume interactions. The last two lines of Eq.~\eqref{free_energy} are associated with the electrostatic interactions between the chain segments, the capsid and the salt ions at the level of Poisson-Boltzmann theory \cite{Siber2008,Borukhov,Shafir}.
The term $W[\Psi]$ represents the free energy density associated with the annealed branching of the polymer \cite{Lubensky,Nguyen-Bruinsma,Lee-Nguyen,Elleuch},
\begin{align} \label{W_branched}
  W[\Psi]&= -\frac{1}{\sqrt{\ell^3}}(f_e\Psi+\frac{\ell^3}{6} f_b \Psi^3),
\end{align}
where $f_e$ and $f_b$ are the fugacities of the end and branched points of the annealed polymer, respectively \cite{adsorption2015}. Note that the stem-loop or hair-pin configurations of RNA are counted as end points. The quantity $\frac{1}{\sqrt{\ell^3}} f_e\Psi$ indicates the density of end points and $\frac{\sqrt{\ell^3}}{6} f_b \Psi^3$ the density of branch points. The number of end $N_e$ and branched points $N_b$ are related to the fugacities $f_e$ and $f_b$, respectively, and can be written as
\begin{align}\label{NeNb}
N_e =- \beta f_e \frac{\partial{F}}{\partial{f_e}} \qquad {\rm and} \qquad N_b =- \beta f_b \frac{\partial{F}}{\partial{f_b}}.
\end{align}
There are two additional constraints in the problem. Note first that the total number of monomers (Kuhn lengths) inside the capsid is fixed \cite{deGennes,Hone},
\begin{equation}\label{constraint}
  N = \int {\mathrm{d}^3}{\bf{r}} \; \Psi^2 ({\bf{r}}).	
\end{equation}
We impose this constraint through a Lagrange multiplier, $\mathcal{E}$, introduced below. Second, there is a relation between the number of the end and branched points,
\begin{equation}\label{branch_constraint}
  N_e = N_b+2,
\end{equation}
as there is only a single polymer in the cavity that by construction has no closed loops as it has to mimic the secondary structure of an RNA. The polymer is linear if $f_b=0$, and the number of branched points increases with \textrm{increasing value of $f_b$}. For our calculations, we vary $f_b$ and find $f_e$ through Eqs.~\eqref{NeNb} and ~\eqref{branch_constraint}. Thus, $f_e$ is not a free parameter.

Varying the free energy functional with respect to the monomer density field $\Psi(r)$ and the electrostatic potential $\Phi(r)$, subject to the constraint that the total number of monomers inside the capsid is constant \cite{Hone}, we obtain two self-consistent non-linear differential equations, which couple the monomer density with the electrostatic potential in the interior of the capsid. The resulting equations are
\begin{subequations} \label{eq:diff}
\begin{align}
&\frac{\ell^2}6\nabla^2\Psi=-\mathcal{E}\Psi({\mathbf r})+ \beta\tau \Phi({\mathbf r})\Psi({\mathbf r})+\upsilon\Psi^3+\frac{1}{2}\frac{\partial{W}}{\partial{\Psi}} \\
&\tfrac{\beta e^2}{4\pi \lambda_B} \nabla^2\Phi_{in}({\mathbf r}) = 2\mu e\sinh \beta e\Phi_{in}({\mathbf r}) \! - \! \tau \Psi^2({\mathbf r}) \\
&\tfrac{\beta e^2}{4\pi \lambda_B} \nabla^2\Phi_{out}({\mathbf r}) = 2\mu e\sinh \beta e\Phi_{out}({\mathbf r}) \!  \label{eq:diff_b}
\end{align}
\end{subequations}
with $\mathcal{E}$ the earlier mentioned Lagrange multiplier enforcing the fixed number of monomers in the cavity. The boundary conditions for the electrostatic potential inside and outside of the spherical shell of radius $R$ are,
\begin{subequations}\label{eq:BCEL}
\begin{align}
{\hat n\cdot}{\nabla \Phi_{in}}\mid_{r=R}-{\hat n\cdot}{\nabla \Phi_{out}}\mid_{r=R}&= {{4\pi\lambda_B}}{\sigma}/\beta e^2\label{eq:BCEL_a}\\
\Phi_{in}({\bf r})\mid_{r=R}&=\Phi_{out}({\bf r})\mid_{r=R}\label{eq:BCEL_b}\\
\Phi_{out}(\bf {r})\mid_{r=\infty}&=0. 
\end{align}
\end{subequations}

The boundary condition (BC) for the electrostatic potential is obtained by minimizing the free energy assuming the surface charge density $\sigma$ is fixed. The concentration of the polymer outside of the capsid is assumed to be zero.  The BC for the inside monomer density field $\Psi$ is of Neumann type (${\hat n\cdot}{\nabla \Psi}|_s=0$) that can be obtained from the energy minimization \cite{Hone} but our findings are robust and our conclusion do not change if we impose the Dirichlet boundary condition $\Psi(r)\mid_{r=R}=0$. The former represent a neutral surface, whilst the latter a repelling surface.\cite{deGennes1979}

\begin{figure}
\centering
\includegraphics[width=0.45\textwidth]{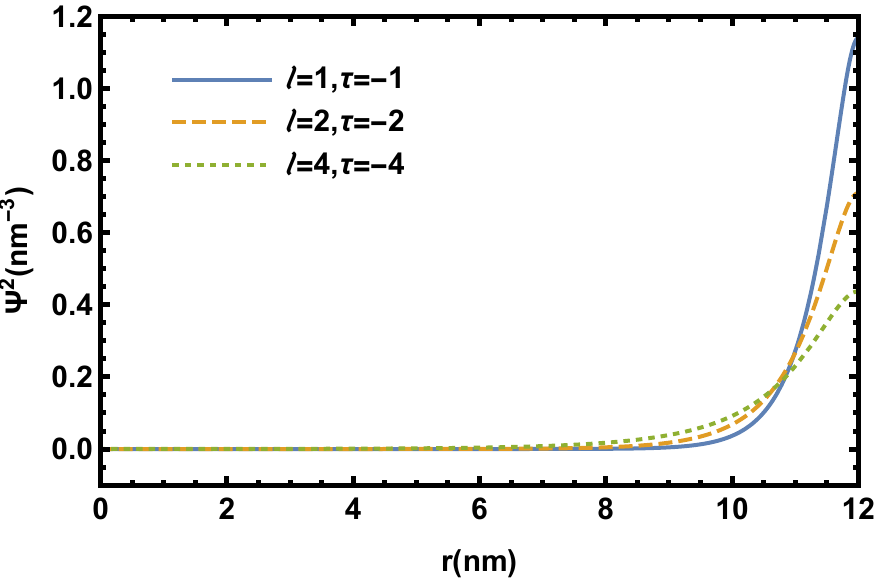}
\caption{\footnotesize Genome density profile as a function of distance from the capsid center for a linear polymer with $l=1 nm$ (solid line),  $l=2 nm$ (dashed line) and $l=4 nm$ (dotted line). Other parameters used correspond to a $T=3$ virus: the total capsid charges on capsid $Q_c = 1800 e$, the strength of excluded volume interaction $\upsilon =0.05 nm^3$, the fugacity $f_b =0$, the quantity $\mu$ corresponds to a salt concentration of $100 mM$, the capsid radius $R =12 nm$, the temperature $T=300 K$ and total number of nucleotides for all three cases equals 1000.
}
\label{profile}
\end{figure}

\section{Results}
We solved the coupled equations given in Eqs.~\eqref{eq:diff} for the $\Psi$ and $\Phi$ fields, subject to the boundary conditions in Eqs.~\eqref{eq:BCEL} through a finite element method (FEM). The polymer density profiles $\Psi^2$ as a function of  the distance from the center of the shell, $r$, are shown in Fig. \ref{profile} for different values of the RNA stiffness $\ell$ and a fixed number of nucleotides, presuming the RNA not to have any secondary structure. Note that for simplicity we assume that a linear chain with $\ell=1$ $nm$ contains one nucleotide and carries one negative charge, so $\tau=-e$. $\ell=2$ $nm$ has two nucleotides with two negative charges and so on. Thus in our figures the numerical value of $\ell$ also indicates the number of nucleotides in one Kuhn length for linear chains. For the three plots in Fig.~\ref{profile}, the total number of nucleotides is calculated using Eq.~\ref{constraint} and is equal to 1000. It is worth mentioning that Eq.~\ref{constraint} gives us the total number of Kuhn lengths $N$ and we multiply it by $\ell$ the number of nucleotides along one Kuhn length to obtain the total number of nucleotides.

As illustrated in the figure, the polymer density becomes larger at the wall as the Kuhn length decreases, even though the linear charge density is fixed.  In all plots for Fig.~\ref{profile} we assumed that the excluded volume is kept constant. Arguably, the excluded volume parameter $\upsilon$ depends on $\ell$, and usually it is assumed that $\upsilon \propto \ell^3$\cite{deGennes1979}.  As we will discuss in Sect.~\ref{discussion}, our conclusions about the role of stiffness in the encapsidation free energy are robust and should not sensitively depend on the strength of the excluded volume interaction.



\begin{figure}
\centering
\includegraphics[width=0.45\textwidth]{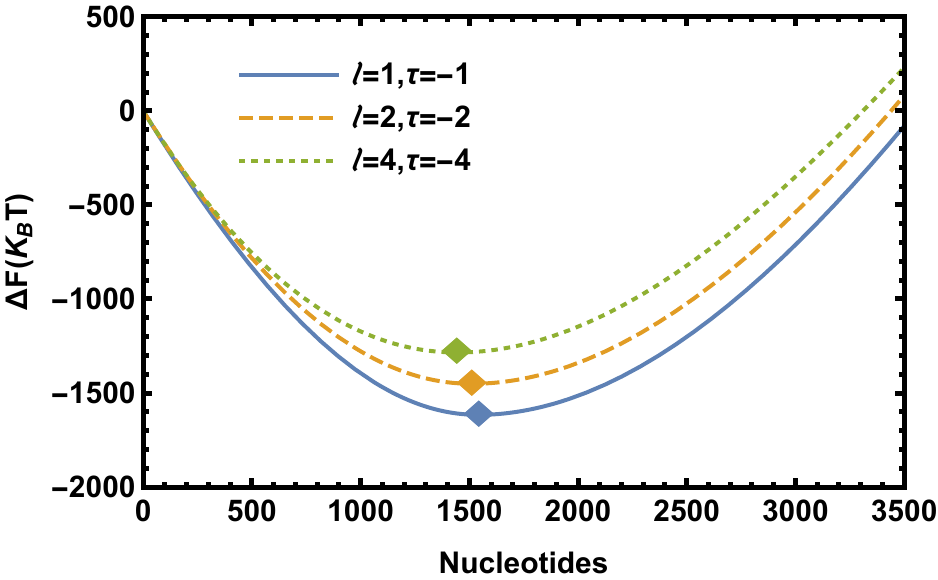}
\caption{\footnotesize Encapsidation free energy of a linear polymer as a function of number of nucleotides for $\ell = 1 nm$ (solid line),  $\ell = 2 nm$ (dashed line) and $\ell = 4 nm$ (dotted line). As the stiffness $\ell$ increases, the optimal number of nucleotides moves towards shorter chains. The quantity $\tau$ indicates the number of negative charges in one Kuhn segment. Other parameters used are the total number of charges on the capsid $Q_c = 1800$, the excluded volume parameter $\upsilon =0.05 nm^3$, the quantity $\mu$ corresponds to a salt concentration of $100 mM$, the radius of the cavity of the capsid $R=12 nm$ and the absolute temperature $T=300 K$.}

\label{Flin}
\end{figure}

To investigate the packaging efficiency of a linear chain as a function of its stiffness, we obtained the free energy of the encapsidation of the linear polymer model as a function of number of nucleotides for different values of $\ell$, as illustrated in Fig.~\ref{Flin}.  The figure shows that the optimal number of nucleotides trapped in the shell increases as $\ell$ decreases. We emphasize again that since we assumed that the size of a single nucleotide is about one $nm$, the numerical value of $\ell$ represents the number of nucleotides within one Kuhn length. This implies that the number of nucleotides and hence the number of charges per Kuhn segment should increase as the Kuhn length increases. For example, in our parametrization $\ell = 4 nm$ represents four nucleotides (resulting in $\tau = - 4 e$).
We observe the same behavior for the free energy of branched polymers, that is, increasing $\ell$ causes the optimal length of genome to move towards shorter chains.  Obviously the stiffness value $\ell$ is larger for the RNAs whose average number of base pairs in the duplex segments is larger.

The concept of the number of nucleotides per Kuhn length is trickier to implement for the branched polymers taken as model for self-hybridized ssRNA. For example, a branched polymer with the Kuhn length $\ell = 1 nm$ represents in our model description two nucleotides and a charge of $\tau = - 2 e$. When the average number of base pairs is about 8 in duplex segments of an ssRNA, we consider the Kuhn length is about eight $nm$, but the number of nucleotides and number of charges per Kuhn length $\tau$ will be 16.  Thus, in our prescription of the self-hybridized ssRNA the number of nucleotides is twice the value of $\ell$ within a Kuhn length as a result of base pairing.

\begin{figure}
\centering
\includegraphics[width=0.45\textwidth]{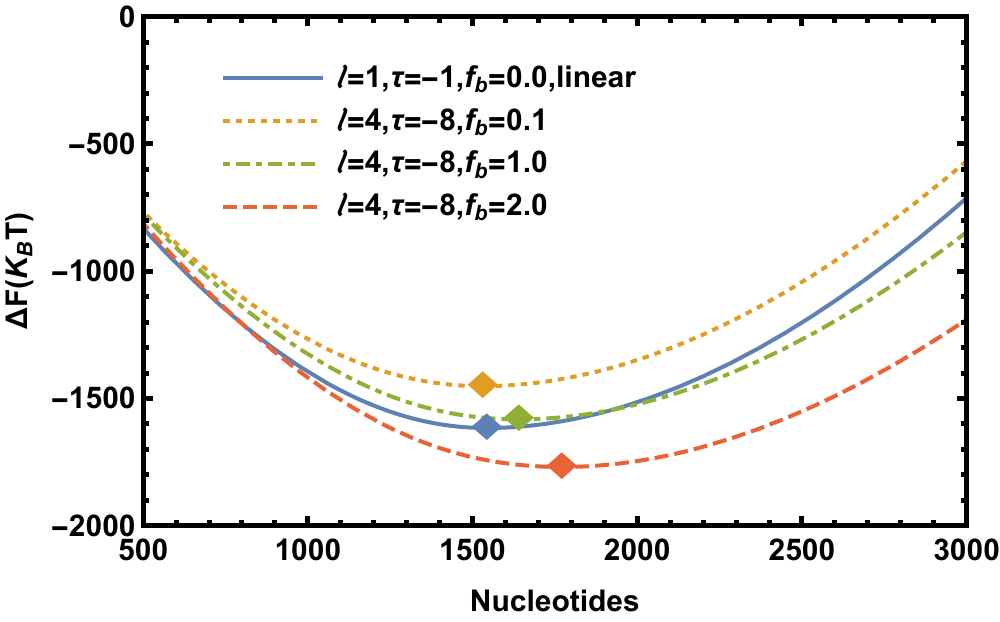}
\caption{\footnotesize  {Encapsidation free energy as a function of number of nucleotides for a linear (solid line) and branched chains with different degree of branching: $f_b=0.1$ (dotted line),  $f_b=1$ (dot-dashed line) and $f_b=2$ (dashed line).} As the fugacity $f_b$ (and hence the number of branched points) increases, the optimal number of nucleotides moves towards longer chains. Other parameters are $Q_c = 1800e$, $\upsilon = 0.05 nm^3$, the quantity $\mu$ corresponds to a salt concentration of $100 mM$, $R=12 nm$ and $T=300 K$.}
\label{bradiffb}
\end{figure}

We also examined the impact of the fugacity on the optimal number of nucleotides. There is a direct relation between the fugacity and the number of branched points: As the fugacity increases the number of branched points of RNA increases too, see \cite{Gonca2014,Gonca2016,Li2017}. Figure~\ref{bradiffb} illustrates that the optimal number of nucleotides increases and the encapsidation free energy becomes more negative, indicating a more stable complex, as the fugacity of branching and hence the number of branch points increases. The solid line in the figure shows the free energy of a linear polymer. For the case shown in the figure, the Kuhn length of the linear chain is $\ell=1 nm$ but that for the branched polymers $\ell=4 nm$, corresponding to four base-paired nucleotides. The number of charges within one Kuhn length then is $\tau=-8 e$.

Figure~\ref{bradiffb} reveals that the free energy of the linear chain is lower than that of the branched one in certain regions of parameter space. For example, for a branched polymer with fugacity $f_b=0.1$, $\ell=4 nm$ and $\tau=-8 e$ (dotted line), the encapsidation free energy of a linear chain with $\ell=1 nm$ and $\tau = -e$ is always lower than that of the branched polymer, and thus, in a head-to-head competition with a limited number of proteins, the linear chain will be the one that is preferentially encapsidated by capsid proteins. This shows that the work of compaction of linear chains could be lower than that of a branched polymer, depending on the stiffness and the degree of branching of the polymers involved. Note that for a fixed $\ell$ while the number of branch points ($f_b$) increases, at some point, the branched polymers outcompetes the linear polymer for binding to capsid proteins, as is illustrated in the figure.

\begin{figure}
\centering
\includegraphics[width=0.45\textwidth]{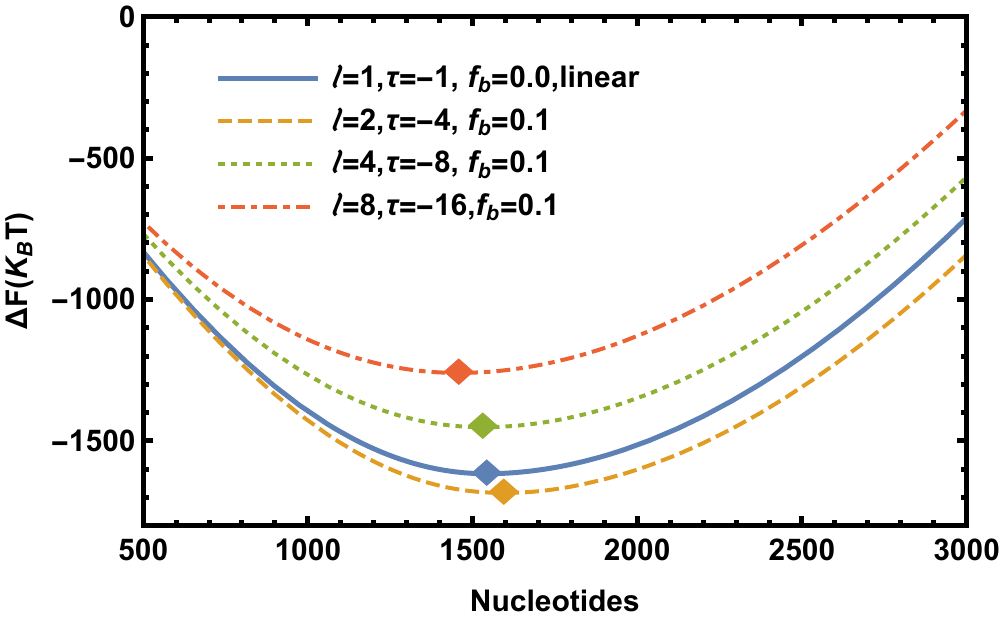}
\caption{\footnotesize  {Encapsidation free energy as a function of number of nucleotides for a linear (solid line) and a branched chain at $\ell=2 nm$ (dashed line),  $\ell=4 nm$ (dotted line) and $\ell=8 nm$ (dot-dashed line).} Other parameters used are $Q_c = 1800 e$, $\upsilon =0.05 nm^3$, the quantity $\mu$ corresponds to a salt concentration of $100 mM$, $R=12 nm$ and $T = 300 K$.}
\label{linwin}
\end{figure}

We next studied the free energy of a branched polymer with a fixed fugacity for different values of the stiffness $\ell$. As illustrated in Fig.~\ref{linwin} for a fugacity $f_b=0.1$, the linear chain (solid) ``looses" to a branched one when four nucleotides have formed two base pairs with $\ell=2 nm$ and $\tau=-4 e$ (dashed line). However, the figure shows that as $\ell$ increases, for $\ell=4 nm$ and $8 nm$ (dotted and dotted-dashed lines), their encapsidation free energies become larger than that of the linear chain, indicating that in a head-to-head competition the linear polymer will be encapsidated. Thus, if the average number of nucleotides in duplex segments increases, it becomes energetically more costly to confine RNA inside the capsid.

%
\section{Discussion}
\label{discussion}


\begin{figure} 
   \centering
   \includegraphics[width=0.45\textwidth]{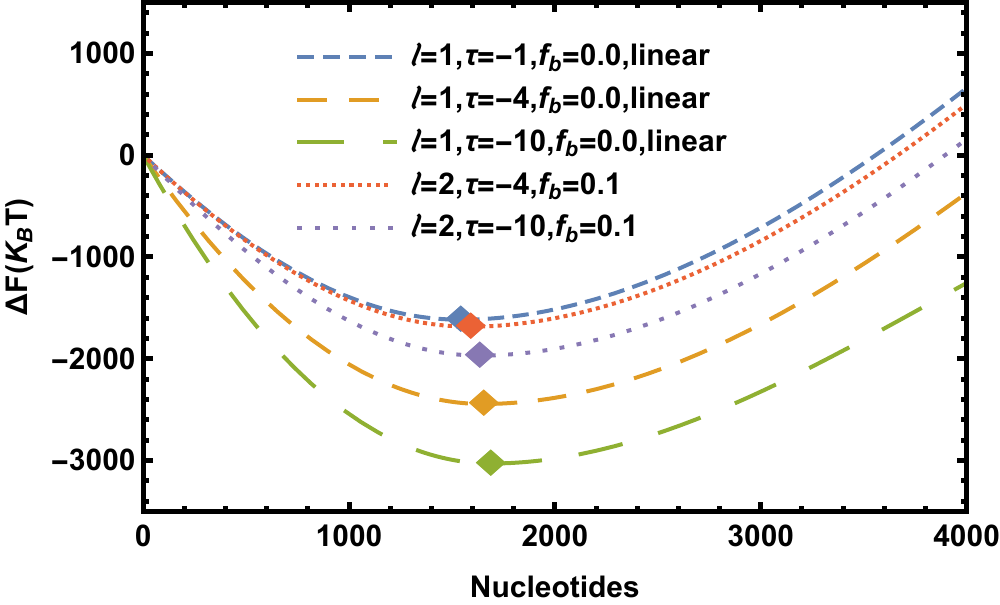}
   \caption{\footnotesize {Encapsidation free energy as a function of the number of nucleotides for linear (dashed lines) and branched chains (dotted lines), with various number of charges within one Kuhn length $\tau$. Other parameters are $Q_c =1800 e$, $\upsilon =0.05 nm^3$, the quantity $\mu$ corresponds to a salt concentration of $100 mM$, $R=12 nm$ and $T=300 K$.}}
   \label{chargedensity}
\end{figure}


Recent experiments emphasized on the crucial role of the RNA topology in the efficiency of virus assembly. As noted in the introduction, Comas-Garcia {\it et al.} \cite{Comas} have shown that CCMV capsid proteins exclusively encapsidate BMV RNA in the presence of the cognate CCMV RNA under conditions where there is a limited number of capsid proteins in solution. The simulations and analytical studies performed in Refs.~\cite{Gonca2014,Gonca2016,Li2017,elife,GoncaPRL2017} are consistent with these results: the viral RNA with a larger degree of branching has a competitive edge over the other viral RNAs or non-viral randomly branched RNAs, keeping all other chain quantities equal.

Indeed, all mean-field theories, numerical calculations and simulations up to now have indicated that the encapsidation free energy of both annealed and quenched branched polymers is significantly lower than that of linear polymers. This suggests that if there are equal amounts of linear and branched polymers in a solution, but there are sufficient capsid proteins to encapsulate exclusively half of the genomes in solutions, only the branched polymer is encapsidated by capsid proteins. Nevertheless, according to a series of more recent experiments by Beren \textit{et al.} \cite{Beren2017} in a head-to head competition between a linear (polyU) chain and CCMV RNA of equal length, surprisingly, and in contrast to theoretical predictions, the linear chain outcompetes the cognate RNA.

While previous theoretical studies have focused on the scaling behavior of linear and branched flexible polymers \cite{Vanderschoot2009,Gonca2014,Gonca2016,Li2017,Siber2008,SiberZandi2010,GoncaPRL2017}, in this paper we study the impact of the stiffness or Kuhn length on the encapsidation of RNA by capsid proteins. In general the duplexed segments of viral RNA contain on average about five to six base-pairs \cite{Fang2011}. Note that some studies show that viral RNAs must have between 60 and 70 per cent of their nucleotides in duplexes, so the linear charge density is almost a factor of two larger and the effective chain length about twice shorter \cite{BORODAVKA2016}.  We argue that while the base pairing on the one hand makes the RNA more compact, on the other hand it increases the effective Kuhn length or the statistical length of the polymer unit. This leads to an increase in the work of compaction of the flexible chain by capsid proteins, which is directly related to the encapsidation free energy of the polymer as plotted in Fig.~\ref{linwin}. We emphasize again that the findings of this paper is not in contradiction with the previous studies: The more strongly branched a polymer is, the more competitive it becomes to be encapsidated by capsid proteins. However, in this work we show that because of base-pairing, the RNA also becomes stiffer and under appropriate conditions can no longer outcompete the linear polymer for binding to capsid proteins.
%

Since branching due to base-pairing causes both the stiffness and the linear charge density of an otherwise linear polymer to increase, one might wonder which effect, higher charge density or larger stiffness, makes the viral RNA less competitive than a linear polymer. Figure \ref{chargedensity} distinguishes the effect of stiffness and charge density. The dashed lines in the figure correspond to linear polymers with $\ell=1 nm$ but different numbers of charges per Kuhn segment $\tau=-e,-4e,-10e$. In the plots, the longer the dashes are, the higher the charge density is. As illustrated in the figure, the encapsidation free energy becomes lower as the charge density increases. The charge density has the same impact on the encapsidation free energy of branched polymers. Figure \ref{chargedensity} shows that as the charge density of branched polymer increases (dotted lines), their free energy decreases. The more distance between the dots, the higher the charge density of the branched polymer.  Quite interestingly, the figure shows that the effect of stiffness overshadows the impact of charge density. A branched polymer with the stiffness of $\ell=2 nm$ and charge density of $\tau=-4e$ or $-10e$ has a higher free energy than a linear polymer with the stiffness of $\ell=1 nm$ but the charge density of $\tau=-4e$. These examples do not correspond to ``real'' RNA as it is not possible to increase the number of charges to more than $2e$ per base pair, but they clarify that base-pairing has three competing effects. First, it makes RNA stiffer, which increases the work of encapsidation but, second, in parallel gives rise to the branching effect and, third, a higher charge density, which both lowers the encapsidation free energy and enhances the packaging efficiency of RNA by capsid proteins.

Another important point to consider, is the change in the excluded volume interaction that must somehow be connected with the variation in the Kuhn length. We repeated the calculations done for Fig. \ref{linwin}, but considered the excluded volume effect, which approximately goes as $\ell^3$ \cite{deGennes1979}. We found that our conclusion is robust and that the excluded volume interaction only slightly modifies the boundary in the parameter space where the linear polymers are able to outcompete the branched ones. The results of this study can explain the intriguing findings of the experiments of Beren \textit{et al.} \cite{Beren2017} in which the unstructured polyU RNA is preferentially packaged and outcompetes native RNA CCMV, despite the fact that viral RNAs have more branch points and as such have a more compact structure. Last but not least we note that the interaction of RNA with capsid proteins could modify the preferred curvature of proteins and result into the capsid of different sizes and $T$ numbers as demonstrated in \cite{Beren2017}.  However, since very little is known about this effect, in this paper we exclusively focused on the impact of RNA stiffness resulting from its base pairing in the RNA encapsidation free energy.

\section{Conclusions}
Results of our field theory calculations have shown that competition between different forms of RNA for encapsulation by virus coat proteins is a complex function of the degree of branching, effective stiffness of the polymer, linear charge density and excluded volume interactions. The conclusion of previous works that the more branched an RNA is on account of its secondary, base-paired structure, the larger the competitive edge it has to be encapsulated in the presence of coat proteins needs to be refined. Under appropriate conditions of linear charge density and effective chain stiffness, we find that a linear chain may in fact outcompete even the native RNA of a virus, as was recently also shown experimentally. Of course, our conclusions are based on coarse-grained model in which the RNA binding domains of the coat proteins are represented by a smooth, uniformly charged wall. In future work we intend to more realistically model these polycationic tails that form a complex with the polynucleotide. Of particular interest here is the impact of excluded volume interactions between these tails and the polynucleotide.
\\
\section*{Acknowledgments}
The authors would like to thank Jef Wagner for useful
discussions. This work was supported by the National Science Foundation through Grant No. DMR -1719550 (R.Z.).


\bibliography{bibfile}

\begin{thebibliography}{59}%
\makeatletter
\providecommand \@ifxundefined [1]{%
 \@ifx{#1\undefined}
}%
\providecommand \@ifnum [1]{%
 \ifnum #1\expandafter \@firstoftwo
 \else \expandafter \@secondoftwo
 \fi
}%
\providecommand \@ifx [1]{%
 \ifx #1\expandafter \@firstoftwo
 \else \expandafter \@secondoftwo
 \fi
}%
\providecommand \natexlab [1]{#1}%
\providecommand \enquote  [1]{``#1''}%
\providecommand \bibnamefont  [1]{#1}%
\providecommand \bibfnamefont [1]{#1}%
\providecommand \citenamefont [1]{#1}%
\providecommand \href@noop [0]{\@secondoftwo}%
\providecommand \href [0]{\begingroup \@sanitize@url \@href}%
\providecommand \@href[1]{\@@startlink{#1}\@@href}%
\providecommand \@@href[1]{\endgroup#1\@@endlink}%
\providecommand \@sanitize@url [0]{\catcode `\\12\catcode `\$12\catcode
  `\&12\catcode `\#12\catcode `\^12\catcode `\_12\catcode `\%12\relax}%
\providecommand \@@startlink[1]{}%
\providecommand \@@endlink[0]{}%
\providecommand \url  [0]{\begingroup\@sanitize@url \@url }%
\providecommand \@url [1]{\endgroup\@href {#1}{\urlprefix }}%
\providecommand \urlprefix  [0]{URL }%
\providecommand \Eprint [0]{\href }%
\providecommand \doibase [0]{http://dx.doi.org/}%
\providecommand \selectlanguage [0]{\@gobble}%
\providecommand \bibinfo  [0]{\@secondoftwo}%
\providecommand \bibfield  [0]{\@secondoftwo}%
\providecommand \translation [1]{[#1]}%
\providecommand \BibitemOpen [0]{}%
\providecommand \bibitemStop [0]{}%
\providecommand \bibitemNoStop [0]{.\EOS\space}%
\providecommand \EOS [0]{\spacefactor3000\relax}%
\providecommand \BibitemShut  [1]{\csname bibitem#1\endcsname}%
\let\auto@bib@innerbib\@empty
\bibitem [{\citenamefont {Higgs}(2000)}]{Higgs2000}%
  \BibitemOpen
  \bibfield  {author} {\bibinfo {author} {\bibfnamefont {P.~G.}\ \bibnamefont
  {Higgs}},\ }\href@noop {} {\bibfield  {journal} {\bibinfo  {journal} {Q. Rev.
  Biophys.}\ }\textbf {\bibinfo {volume} {33}},\ \bibinfo {pages} {199}
  (\bibinfo {year} {2000})}\BibitemShut {NoStop}%
\bibitem [{\citenamefont {Chen}\ \emph {et~al.}(2012)\citenamefont {Chen},
  \citenamefont {Meisburger}, \citenamefont {Pabit}, \citenamefont {Sutton},
  \citenamefont {Webb},\ and\ \citenamefont {Pollack}}]{Chen2011}%
  \BibitemOpen
  \bibfield  {author} {\bibinfo {author} {\bibfnamefont {H.}~\bibnamefont
  {Chen}}, \bibinfo {author} {\bibfnamefont {S.~P.}\ \bibnamefont
  {Meisburger}}, \bibinfo {author} {\bibfnamefont {S.~A.}\ \bibnamefont
  {Pabit}}, \bibinfo {author} {\bibfnamefont {J.~L.}\ \bibnamefont {Sutton}},
  \bibinfo {author} {\bibfnamefont {W.~W.}\ \bibnamefont {Webb}}, \ and\
  \bibinfo {author} {\bibfnamefont {L.}~\bibnamefont {Pollack}},\ }\href
  {\doibase 10.1073/pnas.1119057109} {\bibfield  {journal} {\bibinfo  {journal}
  {PNAS}\ }\textbf {\bibinfo {volume} {109}},\ \bibinfo {pages} {799} (\bibinfo
  {year} {2012})}\BibitemShut {NoStop}%
\bibitem [{\citenamefont {Kebbekus}\ \emph {et~al.}(1995)\citenamefont
  {Kebbekus}, \citenamefont {Draper},\ and\ \citenamefont
  {Hagerman}}]{Kebbekus1995}%
  \BibitemOpen
  \bibfield  {author} {\bibinfo {author} {\bibfnamefont {P.}~\bibnamefont
  {Kebbekus}}, \bibinfo {author} {\bibfnamefont {D.~E.}\ \bibnamefont
  {Draper}}, \ and\ \bibinfo {author} {\bibfnamefont {P.}~\bibnamefont
  {Hagerman}},\ }\href@noop {} {\bibfield  {journal} {\bibinfo  {journal}
  {Biochemistry}\ }\textbf {\bibinfo {volume} {34}},\ \bibinfo {pages} {4354}
  (\bibinfo {year} {1995})}\BibitemShut {NoStop}%
\bibitem [{\citenamefont {Abels}\ \emph {et~al.}(2005)\citenamefont {Abels},
  \citenamefont {Moreno-Herrero}, \citenamefont {van~der Heijden},
  \citenamefont {Dekker},\ and\ \citenamefont {Dekker}}]{Abels2005}%
  \BibitemOpen
  \bibfield  {author} {\bibinfo {author} {\bibfnamefont {J.~A.}\ \bibnamefont
  {Abels}}, \bibinfo {author} {\bibfnamefont {F.}~\bibnamefont
  {Moreno-Herrero}}, \bibinfo {author} {\bibfnamefont {T.}~\bibnamefont
  {van~der Heijden}}, \bibinfo {author} {\bibfnamefont {C.}~\bibnamefont
  {Dekker}}, \ and\ \bibinfo {author} {\bibfnamefont {N.~H.}\ \bibnamefont
  {Dekker}},\ }\href {\doibase 10.1529/biophysj.104.052811} {\bibfield
  {journal} {\bibinfo  {journal} {Biophysical Journal}\ }\textbf {\bibinfo
  {volume} {88}},\ \bibinfo {pages} {2737} (\bibinfo {year}
  {2005})}\BibitemShut {NoStop}%
\bibitem [{\citenamefont {Nussinov}\ \emph {et~al.}(1978)\citenamefont
  {Nussinov}, \citenamefont {Pieczenik}, \citenamefont {Griggs},\ and\
  \citenamefont {Kleitman}}]{Nussinov}%
  \BibitemOpen
  \bibfield  {author} {\bibinfo {author} {\bibfnamefont {R.}~\bibnamefont
  {Nussinov}}, \bibinfo {author} {\bibfnamefont {G.}~\bibnamefont {Pieczenik}},
  \bibinfo {author} {\bibfnamefont {J.~R.}\ \bibnamefont {Griggs}}, \ and\
  \bibinfo {author} {\bibfnamefont {D.~J.}\ \bibnamefont {Kleitman}},\ }\href
  {\doibase 10.1137/0135006} {\bibfield  {journal} {\bibinfo  {journal} {SIAM
  Journal on Applied Mathematics}\ }\textbf {\bibinfo {volume} {35}},\ \bibinfo
  {pages} {68} (\bibinfo {year} {1978})}\BibitemShut {NoStop}%
\bibitem [{\citenamefont {McCaskill}(1990)}]{McCaskill1990}%
  \BibitemOpen
  \bibfield  {author} {\bibinfo {author} {\bibfnamefont {J.~S.}\ \bibnamefont
  {McCaskill}},\ }\href {\doibase 10.1002/bip.360290621} {\bibfield  {journal}
  {\bibinfo  {journal} {Biopolymers}\ }\textbf {\bibinfo {volume} {29}},\
  \bibinfo {pages} {1105} (\bibinfo {year} {1990})}\BibitemShut {NoStop}%
\bibitem [{\citenamefont {Zuker}(2003)}]{Zuker}%
  \BibitemOpen
  \bibfield  {author} {\bibinfo {author} {\bibfnamefont {M.}~\bibnamefont
  {Zuker}},\ }\href {\doibase 10.1093/nar/gkg595} {\bibfield  {journal}
  {\bibinfo  {journal} {Nucleic Acids Res.}\ }\textbf {\bibinfo {volume}
  {31}},\ \bibinfo {pages} {3406} (\bibinfo {year} {2003})}\BibitemShut
  {NoStop}%
\bibitem [{\citenamefont {Hofacker}\ \emph {et~al.}(1994)\citenamefont
  {Hofacker}, \citenamefont {Fontana}, \citenamefont {Stadler}, \citenamefont
  {Bonhoeffer}, \citenamefont {Tacker},\ and\ \citenamefont
  {Schuster}}]{Vienna}%
  \BibitemOpen
  \bibfield  {author} {\bibinfo {author} {\bibfnamefont {I.~L.}\ \bibnamefont
  {Hofacker}}, \bibinfo {author} {\bibfnamefont {W.}~\bibnamefont {Fontana}},
  \bibinfo {author} {\bibfnamefont {P.~F.}\ \bibnamefont {Stadler}}, \bibinfo
  {author} {\bibfnamefont {L.~S.}\ \bibnamefont {Bonhoeffer}}, \bibinfo
  {author} {\bibfnamefont {M.}~\bibnamefont {Tacker}}, \ and\ \bibinfo {author}
  {\bibfnamefont {P.}~\bibnamefont {Schuster}},\ }\href {\doibase
  10.1007/bf00818163} {\bibfield  {journal} {\bibinfo  {journal} {Monatsh.
  Chem.}\ }\textbf {\bibinfo {volume} {125}},\ \bibinfo {pages} {167} (\bibinfo
  {year} {1994})}\BibitemShut {NoStop}%
\bibitem [{\citenamefont {Schwab}\ and\ \citenamefont
  {Bruinsma}(2009)}]{Schwab2009}%
  \BibitemOpen
  \bibfield  {author} {\bibinfo {author} {\bibfnamefont {D.}~\bibnamefont
  {Schwab}}\ and\ \bibinfo {author} {\bibfnamefont {R.~F.}\ \bibnamefont
  {Bruinsma}},\ }\href@noop {} {\bibfield  {journal} {\bibinfo  {journal} {J
  Phys Chem B.}\ }\textbf {\bibinfo {volume} {113}},\ \bibinfo {pages} {3880}
  (\bibinfo {year} {2009})}\BibitemShut {NoStop}%
\bibitem [{\citenamefont {Grosberg}(1997)}]{Grosberg97}%
  \BibitemOpen
  \bibfield  {author} {\bibinfo {author} {\bibfnamefont {A.~Y.}\ \bibnamefont
  {Grosberg}},\ }\href@noop {} {\bibfield  {journal} {\bibinfo  {journal}
  {Physics-Uspekhi}\ }\textbf {\bibinfo {volume} {40}},\ \bibinfo {pages} {125}
  (\bibinfo {year} {1997})}\BibitemShut {NoStop}%
\bibitem [{\citenamefont {Fang}\ \emph {et~al.}(2011)\citenamefont {Fang},
  \citenamefont {Gelbart},\ and\ \citenamefont {Ben-Shaul}}]{Fang2011}%
  \BibitemOpen
  \bibfield  {author} {\bibinfo {author} {\bibfnamefont {L.~T.}\ \bibnamefont
  {Fang}}, \bibinfo {author} {\bibfnamefont {W.~M.}\ \bibnamefont {Gelbart}}, \
  and\ \bibinfo {author} {\bibfnamefont {A.}~\bibnamefont {Ben-Shaul}},\ }\href
  {\doibase 10.1063/1.3652763} {\bibfield  {journal} {\bibinfo  {journal}
  {Journal of Chemical Physics}\ }\textbf {\bibinfo {volume} {135}},\ \bibinfo
  {pages} {155105} (\bibinfo {year} {2011})}\BibitemShut {NoStop}%
\bibitem [{\citenamefont {Singaram}\ \emph {et~al.}(2015)\citenamefont
  {Singaram}, \citenamefont {Garmann}, \citenamefont {Knobler}, \citenamefont
  {Gelbart},\ and\ \citenamefont {Ben-Shaul}}]{Ben-Shaul2015}%
  \BibitemOpen
  \bibfield  {author} {\bibinfo {author} {\bibfnamefont {S.~W.}\ \bibnamefont
  {Singaram}}, \bibinfo {author} {\bibfnamefont {R.~F.}\ \bibnamefont
  {Garmann}}, \bibinfo {author} {\bibfnamefont {C.~M.}\ \bibnamefont
  {Knobler}}, \bibinfo {author} {\bibfnamefont {W.~M.}\ \bibnamefont
  {Gelbart}}, \ and\ \bibinfo {author} {\bibfnamefont {A.}~\bibnamefont
  {Ben-Shaul}},\ }\href@noop {} {\bibfield  {journal} {\bibinfo  {journal} {The
  Journal of Physical Chemistry B}\ }\textbf {\bibinfo {volume} {119}},\
  \bibinfo {pages} {13991} (\bibinfo {year} {2015})}\BibitemShut {NoStop}%
\bibitem [{\citenamefont {Sikkema}\ \emph {et~al.}(2007)\citenamefont
  {Sikkema}, \citenamefont {Comellas-Aragones}, \citenamefont {Fokkink},
  \citenamefont {Verduin}, \citenamefont {Cornelissen},\ and\ \citenamefont
  {Nolte}}]{Cornelissen2007}%
  \BibitemOpen
  \bibfield  {author} {\bibinfo {author} {\bibfnamefont {F.~D.}\ \bibnamefont
  {Sikkema}}, \bibinfo {author} {\bibfnamefont {M.}~\bibnamefont
  {Comellas-Aragones}}, \bibinfo {author} {\bibfnamefont {R.~G.}\ \bibnamefont
  {Fokkink}}, \bibinfo {author} {\bibfnamefont {B.~J.~M.}\ \bibnamefont
  {Verduin}}, \bibinfo {author} {\bibfnamefont {J.}~\bibnamefont
  {Cornelissen}}, \ and\ \bibinfo {author} {\bibfnamefont {R.~J.~M.}\
  \bibnamefont {Nolte}},\ }\href {\doibase 10.1039/b613890j} {\bibfield
  {journal} {\bibinfo  {journal} {Org. Biomol. Chem.}\ }\textbf {\bibinfo
  {volume} {5}},\ \bibinfo {pages} {54} (\bibinfo {year} {2007})}\BibitemShut
  {NoStop}%
\bibitem [{\citenamefont {Ren}\ \emph {et~al.}(2006)\citenamefont {Ren},
  \citenamefont {Wong},\ and\ \citenamefont {Lim}}]{Ren2006}%
  \BibitemOpen
  \bibfield  {author} {\bibinfo {author} {\bibfnamefont {Y.~P.}\ \bibnamefont
  {Ren}}, \bibinfo {author} {\bibfnamefont {S.~M.}\ \bibnamefont {Wong}}, \
  and\ \bibinfo {author} {\bibfnamefont {L.~Y.}\ \bibnamefont {Lim}},\ }\href
  {\doibase 10.1099/vir.0.81944-0} {\bibfield  {journal} {\bibinfo  {journal}
  {J. Gen. Virol.}\ }\textbf {\bibinfo {volume} {87}},\ \bibinfo {pages} {2749}
  (\bibinfo {year} {2006})}\BibitemShut {NoStop}%
\bibitem [{\citenamefont {Ni}\ \emph {et~al.}(2012)\citenamefont {Ni},
  \citenamefont {Wang}, \citenamefont {Ma}, \citenamefont {Das}, \citenamefont
  {Sokol}, \citenamefont {Chiu}, \citenamefont {Dragnea}, \citenamefont
  {Hagan},\ and\ \citenamefont {Kao}}]{Bogdan}%
  \BibitemOpen
  \bibfield  {author} {\bibinfo {author} {\bibfnamefont {P.}~\bibnamefont
  {Ni}}, \bibinfo {author} {\bibfnamefont {Z.}~\bibnamefont {Wang}}, \bibinfo
  {author} {\bibfnamefont {X.}~\bibnamefont {Ma}}, \bibinfo {author}
  {\bibfnamefont {N.~C.}\ \bibnamefont {Das}}, \bibinfo {author} {\bibfnamefont
  {P.}~\bibnamefont {Sokol}}, \bibinfo {author} {\bibfnamefont
  {W.}~\bibnamefont {Chiu}}, \bibinfo {author} {\bibfnamefont {B.}~\bibnamefont
  {Dragnea}}, \bibinfo {author} {\bibfnamefont {M.}~\bibnamefont {Hagan}}, \
  and\ \bibinfo {author} {\bibfnamefont {C.~C.}\ \bibnamefont {Kao}},\ }\href
  {\doibase 10.1016/j.jmb.2012.03.023} {\bibfield  {journal} {\bibinfo
  {journal} {J. Mol. Biol.}\ }\textbf {\bibinfo {volume} {419}},\ \bibinfo
  {pages} {284} (\bibinfo {year} {2012})}\BibitemShut {NoStop}%
\bibitem [{\citenamefont {Losdorfer~Bozic}\ \emph {et~al.}(2013)\citenamefont
  {Losdorfer~Bozic}, \citenamefont {Siber},\ and\ \citenamefont
  {Podgornik}}]{Anze2}%
  \BibitemOpen
  \bibfield  {author} {\bibinfo {author} {\bibfnamefont {A.}~\bibnamefont
  {Losdorfer~Bozic}}, \bibinfo {author} {\bibfnamefont {A.}~\bibnamefont
  {Siber}}, \ and\ \bibinfo {author} {\bibfnamefont {R.}~\bibnamefont
  {Podgornik}},\ }\href@noop {} {\bibfield  {journal} {\bibinfo  {journal} {J.
  Biol. Phys.}\ }\textbf {\bibinfo {volume} {39}},\ \bibinfo {pages} {215}
  (\bibinfo {year} {2013})}\BibitemShut {NoStop}%
\bibitem [{\citenamefont {Zlotnick}\ \emph {et~al.}(2000)\citenamefont
  {Zlotnick}, \citenamefont {Aldrich}, \citenamefont {Johnson}, \citenamefont
  {Ceres},\ and\ \citenamefont {Young}}]{Zlotnick}%
  \BibitemOpen
  \bibfield  {author} {\bibinfo {author} {\bibfnamefont {A.}~\bibnamefont
  {Zlotnick}}, \bibinfo {author} {\bibfnamefont {R.}~\bibnamefont {Aldrich}},
  \bibinfo {author} {\bibfnamefont {J.~M.}\ \bibnamefont {Johnson}}, \bibinfo
  {author} {\bibfnamefont {P.}~\bibnamefont {Ceres}}, \ and\ \bibinfo {author}
  {\bibfnamefont {M.~J.}\ \bibnamefont {Young}},\ }\href {\doibase
  10.1006/viro.2000.0619} {\bibfield  {journal} {\bibinfo  {journal}
  {Virology}\ }\textbf {\bibinfo {volume} {277}},\ \bibinfo {pages} {450}
  (\bibinfo {year} {2000})}\BibitemShut {NoStop}%
\bibitem [{\citenamefont {Sun}\ \emph {et~al.}(2007)\citenamefont {Sun},
  \citenamefont {DuFort}, \citenamefont {Daniel}, \citenamefont {Murali},
  \citenamefont {Chen}, \citenamefont {Gopinath}, \citenamefont {Stein},
  \citenamefont {De}, \citenamefont {Rotello}, \citenamefont {Holzenburg},
  \citenamefont {Kao},\ and\ \citenamefont {Dragnea}}]{Sun2007}%
  \BibitemOpen
  \bibfield  {author} {\bibinfo {author} {\bibfnamefont {J.}~\bibnamefont
  {Sun}}, \bibinfo {author} {\bibfnamefont {C.}~\bibnamefont {DuFort}},
  \bibinfo {author} {\bibfnamefont {M.-C.}\ \bibnamefont {Daniel}}, \bibinfo
  {author} {\bibfnamefont {A.}~\bibnamefont {Murali}}, \bibinfo {author}
  {\bibfnamefont {C.}~\bibnamefont {Chen}}, \bibinfo {author} {\bibfnamefont
  {K.}~\bibnamefont {Gopinath}}, \bibinfo {author} {\bibfnamefont
  {B.}~\bibnamefont {Stein}}, \bibinfo {author} {\bibfnamefont
  {M.}~\bibnamefont {De}}, \bibinfo {author} {\bibfnamefont {V.~M.}\
  \bibnamefont {Rotello}}, \bibinfo {author} {\bibfnamefont {A.}~\bibnamefont
  {Holzenburg}}, \bibinfo {author} {\bibfnamefont {C.~C.}\ \bibnamefont {Kao}},
  \ and\ \bibinfo {author} {\bibfnamefont {B.}~\bibnamefont {Dragnea}},\
  }\href@noop {} {\bibfield  {journal} {\bibinfo  {journal} {Proc. Nat. Acad.
  Sci. USA}\ }\textbf {\bibinfo {volume} {104}},\ \bibinfo {pages} {1354}
  (\bibinfo {year} {2007})}\BibitemShut {NoStop}%
\bibitem [{\citenamefont {Ning}\ \emph {et~al.}(2016)\citenamefont {Ning},
  \citenamefont {Erdemci-Tandogan}, \citenamefont {Yufenyuy}, \citenamefont
  {Wagner}, \citenamefont {Himes}, \citenamefont {Zhao}, \citenamefont {Aiken},
  \citenamefont {Zandi},\ and\ \citenamefont {Zhang}}]{Nature2016}%
  \BibitemOpen
  \bibfield  {author} {\bibinfo {author} {\bibfnamefont {J.}~\bibnamefont
  {Ning}}, \bibinfo {author} {\bibfnamefont {G.}~\bibnamefont
  {Erdemci-Tandogan}}, \bibinfo {author} {\bibfnamefont {E.~L.}\ \bibnamefont
  {Yufenyuy}}, \bibinfo {author} {\bibfnamefont {J.}~\bibnamefont {Wagner}},
  \bibinfo {author} {\bibfnamefont {B.~A.}\ \bibnamefont {Himes}}, \bibinfo
  {author} {\bibfnamefont {G.}~\bibnamefont {Zhao}}, \bibinfo {author}
  {\bibfnamefont {C.}~\bibnamefont {Aiken}}, \bibinfo {author} {\bibfnamefont
  {R.}~\bibnamefont {Zandi}}, \ and\ \bibinfo {author} {\bibfnamefont
  {P.}~\bibnamefont {Zhang}},\ }\href {\doibase 10.1038/ncomms13689} {\bibfield
   {journal} {\bibinfo  {journal} {Nature Communications}\ }\textbf {\bibinfo
  {volume} {7}},\ \bibinfo {pages} {13689} (\bibinfo {year}
  {2016})}\BibitemShut {NoStop}%
\bibitem [{\citenamefont {Fejer}\ \emph {et~al.}(2010)\citenamefont {Fejer},
  \citenamefont {Chakrabarti},\ and\ \citenamefont {Wales}}]{Fejer:10}%
  \BibitemOpen
  \bibfield  {author} {\bibinfo {author} {\bibfnamefont {S.}~\bibnamefont
  {Fejer}}, \bibinfo {author} {\bibfnamefont {D.}~\bibnamefont {Chakrabarti}},
  \ and\ \bibinfo {author} {\bibfnamefont {D.}~\bibnamefont {Wales}},\
  }\href@noop {} {\bibfield  {journal} {\bibinfo  {journal} {Nano Lett}\
  }\textbf {\bibinfo {volume} {4}},\ \bibinfo {pages} {219} (\bibinfo {year}
  {2010})}\BibitemShut {NoStop}%
\bibitem [{\citenamefont {Rapaport}(2004)}]{Rapaport:04a}%
  \BibitemOpen
  \bibfield  {author} {\bibinfo {author} {\bibfnamefont {D.~C.}\ \bibnamefont
  {Rapaport}},\ }\href {\doibase 10.1103/PhysRevE.70.051905} {\bibfield
  {journal} {\bibinfo  {journal} {Phys Rev E}\ }\textbf {\bibinfo {volume}
  {70}},\ \bibinfo {pages} {051905} (\bibinfo {year} {2004})}\BibitemShut
  {NoStop}%
\bibitem [{\citenamefont {Wagner}\ and\ \citenamefont
  {Zandi}(2015)}]{Wagner2015956}%
  \BibitemOpen
  \bibfield  {author} {\bibinfo {author} {\bibfnamefont {J.}~\bibnamefont
  {Wagner}}\ and\ \bibinfo {author} {\bibfnamefont {R.}~\bibnamefont {Zandi}},\
  }\href {\doibase http://dx.doi.org/10.1016/j.bpj.2015.07.041} {\bibfield
  {journal} {\bibinfo  {journal} {Biophysical Journal}\ }\textbf {\bibinfo
  {volume} {109}},\ \bibinfo {pages} {956 } (\bibinfo {year}
  {2015})}\BibitemShut {NoStop}%
\bibitem [{\citenamefont {Luque}\ \emph {et~al.}(2010)\citenamefont {Luque},
  \citenamefont {Zandi},\ and\ \citenamefont {Reguera}}]{Luque:2010a}%
  \BibitemOpen
  \bibfield  {author} {\bibinfo {author} {\bibfnamefont {A.}~\bibnamefont
  {Luque}}, \bibinfo {author} {\bibfnamefont {R.}~\bibnamefont {Zandi}}, \ and\
  \bibinfo {author} {\bibfnamefont {D.}~\bibnamefont {Reguera}},\ }\href
  {\doibase 10.1073/pnas.0915122107} {\bibfield  {journal} {\bibinfo  {journal}
  {PNAS}\ }\textbf {\bibinfo {volume} {107}},\ \bibinfo {pages} {5323}
  (\bibinfo {year} {2010})}\BibitemShut {NoStop}%
\bibitem [{\citenamefont {Chen}\ \emph {et~al.}(2007)\citenamefont {Chen},
  \citenamefont {Zhang},\ and\ \citenamefont {Glotzer}}]{Chen:2007b}%
  \BibitemOpen
  \bibfield  {author} {\bibinfo {author} {\bibfnamefont {T.}~\bibnamefont
  {Chen}}, \bibinfo {author} {\bibfnamefont {Z.}~\bibnamefont {Zhang}}, \ and\
  \bibinfo {author} {\bibfnamefont {S.~C.}\ \bibnamefont {Glotzer}},\
  }\href@noop {} {\bibfield  {journal} {\bibinfo  {journal} {PNAS}\ }\textbf
  {\bibinfo {volume} {104}},\ \bibinfo {pages} {717} (\bibinfo {year}
  {2007})}\BibitemShut {NoStop}%
\bibitem [{\citenamefont {Paquay}\ \emph {et~al.}(2016)\citenamefont {Paquay},
  \citenamefont {Kusumaatmaja}, \citenamefont {Wales}, \citenamefont {Zandi},\
  and\ \citenamefont {van~der Schoot}}]{Stefan}%
  \BibitemOpen
  \bibfield  {author} {\bibinfo {author} {\bibfnamefont {S.}~\bibnamefont
  {Paquay}}, \bibinfo {author} {\bibfnamefont {H.}~\bibnamefont
  {Kusumaatmaja}}, \bibinfo {author} {\bibfnamefont {D.~J.}\ \bibnamefont
  {Wales}}, \bibinfo {author} {\bibfnamefont {R.}~\bibnamefont {Zandi}}, \ and\
  \bibinfo {author} {\bibfnamefont {P.}~\bibnamefont {van~der Schoot}},\ }\href
  {\doibase 10.1039/C6SM00489J} {\bibfield  {journal} {\bibinfo  {journal}
  {Soft Matter}\ }\textbf {\bibinfo {volume} {12}},\ \bibinfo {pages} {5708}
  (\bibinfo {year} {2016})}\BibitemShut {NoStop}%
\bibitem [{\citenamefont {Kusters}\ \emph {et~al.}(2015)\citenamefont
  {Kusters}, \citenamefont {Lin}, \citenamefont {Zandi}, \citenamefont
  {Tsvetkova}, \citenamefont {Dragnea},\ and\ \citenamefont {van~der
  Schoot}}]{Kusters2015}%
  \BibitemOpen
  \bibfield  {author} {\bibinfo {author} {\bibfnamefont {R.}~\bibnamefont
  {Kusters}}, \bibinfo {author} {\bibfnamefont {H.-K.}\ \bibnamefont {Lin}},
  \bibinfo {author} {\bibfnamefont {R.}~\bibnamefont {Zandi}}, \bibinfo
  {author} {\bibfnamefont {I.}~\bibnamefont {Tsvetkova}}, \bibinfo {author}
  {\bibfnamefont {B.}~\bibnamefont {Dragnea}}, \ and\ \bibinfo {author}
  {\bibfnamefont {P.}~\bibnamefont {van~der Schoot}},\ }\href {\doibase
  10.1021/jp5108125} {\bibfield  {journal} {\bibinfo  {journal} {J. Phys. Chem.
  B}\ }\textbf {\bibinfo {volume} {119}},\ \bibinfo {pages} {1869} (\bibinfo
  {year} {2015})}\BibitemShut {NoStop}%
\bibitem [{\citenamefont {Hagan}\ and\ \citenamefont
  {Zandi}(2016)}]{Zandi2016}%
  \BibitemOpen
  \bibfield  {author} {\bibinfo {author} {\bibfnamefont {M.~F.}\ \bibnamefont
  {Hagan}}\ and\ \bibinfo {author} {\bibfnamefont {R.}~\bibnamefont {Zandi}},\
  }\href {\doibase 10.1016/j.coviro.2016.02.012} {\bibfield  {journal}
  {\bibinfo  {journal} {Curr. Opin. Virol.}\ }\textbf {\bibinfo {volume}
  {18}},\ \bibinfo {pages} {36} (\bibinfo {year} {2016})}\BibitemShut {NoStop}%
\bibitem [{\citenamefont {Lin}\ \emph {et~al.}(2012)\citenamefont {Lin},
  \citenamefont {van~der Schoot},\ and\ \citenamefont {Zandi}}]{Hsiang-Ku}%
  \BibitemOpen
  \bibfield  {author} {\bibinfo {author} {\bibfnamefont {H.-K.}\ \bibnamefont
  {Lin}}, \bibinfo {author} {\bibfnamefont {P.}~\bibnamefont {van~der Schoot}},
  \ and\ \bibinfo {author} {\bibfnamefont {R.}~\bibnamefont {Zandi}},\ }\href
  {\doibase 10.1088/1478-3975/9/6/066004} {\bibfield  {journal} {\bibinfo
  {journal} {Phys. Biol.}\ }\textbf {\bibinfo {volume} {9}},\ \bibinfo {pages}
  {066004} (\bibinfo {year} {2012})}\BibitemShut {NoStop}%
\bibitem [{\citenamefont {Sivanandam}\ \emph {et~al.}(2016)\citenamefont
  {Sivanandam}, \citenamefont {Mathews}, \citenamefont {Garmann}, \citenamefont
  {Erdemci-Tandogan}, \citenamefont {Zandi},\ and\ \citenamefont
  {Rao}}]{Venky2016}%
  \BibitemOpen
  \bibfield  {author} {\bibinfo {author} {\bibfnamefont {V.}~\bibnamefont
  {Sivanandam}}, \bibinfo {author} {\bibfnamefont {D.}~\bibnamefont {Mathews}},
  \bibinfo {author} {\bibfnamefont {R.}~\bibnamefont {Garmann}}, \bibinfo
  {author} {\bibfnamefont {G.}~\bibnamefont {Erdemci-Tandogan}}, \bibinfo
  {author} {\bibfnamefont {R.}~\bibnamefont {Zandi}}, \ and\ \bibinfo {author}
  {\bibfnamefont {A.~L.~N.}\ \bibnamefont {Rao}},\ }\href
  {http://dx.doi.org/10.1038/srep26328 http://10.1038/srep26328} {\bibfield
  {journal} {\bibinfo  {journal} {Scientific Reports}\ }\textbf {\bibinfo
  {volume} {6}},\ \bibinfo {pages} {26328} (\bibinfo {year}
  {2016})}\BibitemShut {NoStop}%
\bibitem [{\citenamefont {Patel}\ \emph {et~al.}(2015)\citenamefont {Patel},
  \citenamefont {Dykeman}, \citenamefont {Coutts}, \citenamefont {Lomonossoff},
  \citenamefont {Rowlands}, \citenamefont {Phillips}, \citenamefont {Ranson},
  \citenamefont {Twarock}, \citenamefont {Tuma},\ and\ \citenamefont
  {Stockley}}]{Patel2014}%
  \BibitemOpen
  \bibfield  {author} {\bibinfo {author} {\bibfnamefont {N.}~\bibnamefont
  {Patel}}, \bibinfo {author} {\bibfnamefont {E.~C.}\ \bibnamefont {Dykeman}},
  \bibinfo {author} {\bibfnamefont {R.~H.~A.}\ \bibnamefont {Coutts}}, \bibinfo
  {author} {\bibfnamefont {G.~P.}\ \bibnamefont {Lomonossoff}}, \bibinfo
  {author} {\bibfnamefont {D.~J.}\ \bibnamefont {Rowlands}}, \bibinfo {author}
  {\bibfnamefont {S.~E.~V.}\ \bibnamefont {Phillips}}, \bibinfo {author}
  {\bibfnamefont {N.}~\bibnamefont {Ranson}}, \bibinfo {author} {\bibfnamefont
  {R.}~\bibnamefont {Twarock}}, \bibinfo {author} {\bibfnamefont
  {R.}~\bibnamefont {Tuma}}, \ and\ \bibinfo {author} {\bibfnamefont {P.~G.}\
  \bibnamefont {Stockley}},\ }\href {\doibase 10.1073/pnas.1420812112}
  {\bibfield  {journal} {\bibinfo  {journal} {Proceedings of the National
  Academy of Sciences}\ }\textbf {\bibinfo {volume} {112}},\ \bibinfo {pages}
  {2227} (\bibinfo {year} {2015})},\ \Eprint
  {http://arxiv.org/abs/http://www.pnas.org/content/112/7/2227.full.pdf}
  {http://www.pnas.org/content/112/7/2227.full.pdf} \BibitemShut {NoStop}%
\bibitem [{\citenamefont {Comas-Garcia}\ \emph {et~al.}(2012)\citenamefont
  {Comas-Garcia}, \citenamefont {Cadena-Nava}, \citenamefont {Rao},
  \citenamefont {Knobler},\ and\ \citenamefont {Gelbart}}]{Comas}%
  \BibitemOpen
  \bibfield  {author} {\bibinfo {author} {\bibfnamefont {M.}~\bibnamefont
  {Comas-Garcia}}, \bibinfo {author} {\bibfnamefont {R.~D.}\ \bibnamefont
  {Cadena-Nava}}, \bibinfo {author} {\bibfnamefont {A.~L.~N.}\ \bibnamefont
  {Rao}}, \bibinfo {author} {\bibfnamefont {C.~M.}\ \bibnamefont {Knobler}}, \
  and\ \bibinfo {author} {\bibfnamefont {W.~M.}\ \bibnamefont {Gelbart}},\
  }\href {\doibase 10.1128/jvi.01695-12} {\bibfield  {journal} {\bibinfo
  {journal} {J. Virol.}\ }\textbf {\bibinfo {volume} {86}},\ \bibinfo {pages}
  {12271} (\bibinfo {year} {2012})}\BibitemShut {NoStop}%
\bibitem [{\citenamefont {Erdemci-Tandogan}\ \emph {et~al.}(2014)\citenamefont
  {Erdemci-Tandogan}, \citenamefont {Wagner}, \citenamefont {van~der Schoot},
  \citenamefont {Podgornik},\ and\ \citenamefont {Zandi}}]{Gonca2014}%
  \BibitemOpen
  \bibfield  {author} {\bibinfo {author} {\bibfnamefont {G.}~\bibnamefont
  {Erdemci-Tandogan}}, \bibinfo {author} {\bibfnamefont {J.}~\bibnamefont
  {Wagner}}, \bibinfo {author} {\bibfnamefont {P.}~\bibnamefont {van~der
  Schoot}}, \bibinfo {author} {\bibfnamefont {R.}~\bibnamefont {Podgornik}}, \
  and\ \bibinfo {author} {\bibfnamefont {R.}~\bibnamefont {Zandi}},\ }\href
  {\doibase 10.1103/PhysRevE.89.032707} {\bibfield  {journal} {\bibinfo
  {journal} {Phys. Rev. E}\ }\textbf {\bibinfo {volume} {89}},\ \bibinfo
  {pages} {032707} (\bibinfo {year} {2014})}\BibitemShut {NoStop}%
\bibitem [{\citenamefont {Perlmutter}\ \emph {et~al.}(2013)\citenamefont
  {Perlmutter}, \citenamefont {Qiao},\ and\ \citenamefont {Hagan}}]{elife}%
  \BibitemOpen
  \bibfield  {author} {\bibinfo {author} {\bibfnamefont {J.~D.}\ \bibnamefont
  {Perlmutter}}, \bibinfo {author} {\bibfnamefont {C.}~\bibnamefont {Qiao}}, \
  and\ \bibinfo {author} {\bibfnamefont {M.~F.}\ \bibnamefont {Hagan}},\ }\href
  {\doibase 10.7554/eLife.00632} {\bibfield  {journal} {\bibinfo  {journal}
  {eLife}\ }\textbf {\bibinfo {volume} {2}},\ \bibinfo {pages} {e00632}
  (\bibinfo {year} {2013})}\BibitemShut {NoStop}%
\bibitem [{\citenamefont {Erdemci-Tandogan}\ \emph
  {et~al.}(2016{\natexlab{a}})\citenamefont {Erdemci-Tandogan}, \citenamefont
  {Wagner}, \citenamefont {van~der Schoot}, \citenamefont {Podgornik},\ and\
  \citenamefont {Zandi}}]{Gonca2016}%
  \BibitemOpen
  \bibfield  {author} {\bibinfo {author} {\bibfnamefont {G.}~\bibnamefont
  {Erdemci-Tandogan}}, \bibinfo {author} {\bibfnamefont {J.}~\bibnamefont
  {Wagner}}, \bibinfo {author} {\bibfnamefont {P.}~\bibnamefont {van~der
  Schoot}}, \bibinfo {author} {\bibfnamefont {R.}~\bibnamefont {Podgornik}}, \
  and\ \bibinfo {author} {\bibfnamefont {R.}~\bibnamefont {Zandi}},\
  }\href@noop {} {\bibfield  {journal} {\bibinfo  {journal} {Phys. Rev. E}\
  }\textbf {\bibinfo {volume} {94}},\ \bibinfo {pages} {022408} (\bibinfo
  {year} {2016}{\natexlab{a}})}\BibitemShut {NoStop}%
\bibitem [{\citenamefont {Li}\ \emph {et~al.}(2017)\citenamefont {Li},
  \citenamefont {Erdemci-Tandogan}, \citenamefont {Wagner}, \citenamefont {{Van
  Der Schoot}},\ and\ \citenamefont {Zandi}}]{Li2017}%
  \BibitemOpen
  \bibfield  {author} {\bibinfo {author} {\bibfnamefont {S.}~\bibnamefont
  {Li}}, \bibinfo {author} {\bibfnamefont {G.}~\bibnamefont
  {Erdemci-Tandogan}}, \bibinfo {author} {\bibfnamefont {J.}~\bibnamefont
  {Wagner}}, \bibinfo {author} {\bibfnamefont {P.}~\bibnamefont {{Van Der
  Schoot}}}, \ and\ \bibinfo {author} {\bibfnamefont {R.}~\bibnamefont
  {Zandi}},\ }\href {\doibase 10.1103/PhysRevE.96.022401} {\bibfield  {journal}
  {\bibinfo  {journal} {Physical Review E}\ }\textbf {\bibinfo {volume} {96}},\
  \bibinfo {pages} {022401} (\bibinfo {year} {2017})}\BibitemShut {NoStop}%
\bibitem [{\citenamefont {Yoffe}\ \emph {et~al.}(2008)\citenamefont {Yoffe},
  \citenamefont {Prinsen}, \citenamefont {Gopal}, \citenamefont {Knobler},
  \citenamefont {Gelbart},\ and\ \citenamefont {Ben-Shaul}}]{Yoffe2008}%
  \BibitemOpen
  \bibfield  {author} {\bibinfo {author} {\bibfnamefont {A.~M.}\ \bibnamefont
  {Yoffe}}, \bibinfo {author} {\bibfnamefont {P.}~\bibnamefont {Prinsen}},
  \bibinfo {author} {\bibfnamefont {A.}~\bibnamefont {Gopal}}, \bibinfo
  {author} {\bibfnamefont {C.~M.}\ \bibnamefont {Knobler}}, \bibinfo {author}
  {\bibfnamefont {W.~M.}\ \bibnamefont {Gelbart}}, \ and\ \bibinfo {author}
  {\bibfnamefont {A.}~\bibnamefont {Ben-Shaul}},\ }\href {\doibase
  10.1073/pnas.0808089105} {\bibfield  {journal} {\bibinfo  {journal} {PNAS}\
  }\textbf {\bibinfo {volume} {105}},\ \bibinfo {pages} {16153} (\bibinfo
  {year} {2008})}\BibitemShut {NoStop}%
\bibitem [{\citenamefont {Bruinsma}\ \emph {et~al.}(2016)\citenamefont
  {Bruinsma}, \citenamefont {Comas-Garcia}, \citenamefont {Garmann},\ and\
  \citenamefont {Grosberg}}]{Bruinsma2016}%
  \BibitemOpen
  \bibfield  {author} {\bibinfo {author} {\bibfnamefont {R.~F.}\ \bibnamefont
  {Bruinsma}}, \bibinfo {author} {\bibfnamefont {M.}~\bibnamefont
  {Comas-Garcia}}, \bibinfo {author} {\bibfnamefont {R.~F.}\ \bibnamefont
  {Garmann}}, \ and\ \bibinfo {author} {\bibfnamefont {A.~Y.}\ \bibnamefont
  {Grosberg}},\ }\href {\doibase 10.1103/PhysRevE.93.032405} {\bibfield
  {journal} {\bibinfo  {journal} {Physical Review E}\ }\textbf {\bibinfo
  {volume} {93}},\ \bibinfo {pages} {1} (\bibinfo {year} {2016})},\ \Eprint
  {http://arxiv.org/abs/arXiv:1505.01224v1} {arXiv:arXiv:1505.01224v1}
  \BibitemShut {NoStop}%
\bibitem [{\citenamefont {Li~Tai}\ \emph {et~al.}(2011)\citenamefont {Li~Tai},
  \citenamefont {Gelbart},\ and\ \citenamefont {Ben-Shaul}}]{Li-tai}%
  \BibitemOpen
  \bibfield  {author} {\bibinfo {author} {\bibfnamefont {F.}~\bibnamefont
  {Li~Tai}}, \bibinfo {author} {\bibfnamefont {W.~M.}\ \bibnamefont {Gelbart}},
  \ and\ \bibinfo {author} {\bibfnamefont {A.}~\bibnamefont {Ben-Shaul}},\
  }\href {\doibase 10.1063/1.3652763} {\bibfield  {journal} {\bibinfo
  {journal} {J. Chem. Phys.}\ }\textbf {\bibinfo {volume} {135}},\ \bibinfo
  {pages} {155105} (\bibinfo {year} {2011})}\BibitemShut {NoStop}%
\bibitem [{\citenamefont {Gopal}\ \emph {et~al.}(2014)\citenamefont {Gopal},
  \citenamefont {D.E.}, \citenamefont {A.M.}, \citenamefont {A}, \citenamefont
  {ALN}, \citenamefont {Knobler}, \citenamefont {Gelbart},\ and\ \citenamefont
  {Ben-Shaul}}]{Gopal2014}%
  \BibitemOpen
  \bibfield  {author} {\bibinfo {author} {\bibfnamefont {A.}~\bibnamefont
  {Gopal}}, \bibinfo {author} {\bibfnamefont {E.}~\bibnamefont {D.E.}},
  \bibinfo {author} {\bibfnamefont {Y.}~\bibnamefont {A.M.}}, \bibinfo {author}
  {\bibfnamefont {B.-S.}\ \bibnamefont {A}}, \bibinfo {author} {\bibfnamefont
  {R.}~\bibnamefont {ALN}}, \bibinfo {author} {\bibfnamefont {C.~M.}\
  \bibnamefont {Knobler}}, \bibinfo {author} {\bibfnamefont {W.~M.}\
  \bibnamefont {Gelbart}}, \ and\ \bibinfo {author} {\bibfnamefont
  {A.}~\bibnamefont {Ben-Shaul}},\ }\href@noop {} {\bibfield  {journal}
  {\bibinfo  {journal} {PLoS ONE}\ }\textbf {\bibinfo {volume} {9}},\ \bibinfo
  {pages} {e105875} (\bibinfo {year} {2014})}\BibitemShut {NoStop}%
\bibitem [{\citenamefont {Beren}\ \emph {et~al.}(2017)\citenamefont {Beren},
  \citenamefont {Dreesens}, \citenamefont {Liu}, \citenamefont {Knobler},\ and\
  \citenamefont {Gelbart}}]{Beren2017}%
  \BibitemOpen
  \bibfield  {author} {\bibinfo {author} {\bibfnamefont {C.}~\bibnamefont
  {Beren}}, \bibinfo {author} {\bibfnamefont {L.~L.}\ \bibnamefont {Dreesens}},
  \bibinfo {author} {\bibfnamefont {K.~N.}\ \bibnamefont {Liu}}, \bibinfo
  {author} {\bibfnamefont {C.~M.}\ \bibnamefont {Knobler}}, \ and\ \bibinfo
  {author} {\bibfnamefont {W.~M.}\ \bibnamefont {Gelbart}},\ }\href {\doibase
  10.1016/j.bpj.2017.06.038} {\bibfield  {journal} {\bibinfo  {journal}
  {Biophysical Journal}\ }\textbf {\bibinfo {volume} {113}},\ \bibinfo {pages}
  {339} (\bibinfo {year} {2017})}\BibitemShut {NoStop}%
\bibitem [{\citenamefont {van~der Schoot}\ and\ \citenamefont
  {Zandi}(2013)}]{Paul:13a}%
  \BibitemOpen
  \bibfield  {author} {\bibinfo {author} {\bibfnamefont {P.}~\bibnamefont
  {van~der Schoot}}\ and\ \bibinfo {author} {\bibfnamefont {R.}~\bibnamefont
  {Zandi}},\ }\href {\doibase 10.1007/s10867-013-9307-y} {\bibfield  {journal}
  {\bibinfo  {journal} {J. Biol. Phys.}\ }\textbf {\bibinfo {volume} {39}},\
  \bibinfo {pages} {289} (\bibinfo {year} {2013})}\BibitemShut {NoStop}%
\bibitem [{\citenamefont {McPherson}(2005)}]{McPherson}%
  \BibitemOpen
  \bibfield  {author} {\bibinfo {author} {\bibfnamefont {A.}~\bibnamefont
  {McPherson}},\ }\href {\doibase 10.1002/bies.20196} {\bibfield  {journal}
  {\bibinfo  {journal} {BioEssays}\ }\textbf {\bibinfo {volume} {27}},\
  \bibinfo {pages} {447} (\bibinfo {year} {2005})}\BibitemShut {NoStop}%
\bibitem [{\citenamefont {Borukhov}\ \emph {et~al.}(1998)\citenamefont
  {Borukhov}, \citenamefont {Andelman},\ and\ \citenamefont
  {Orland}}]{Borukhov}%
  \BibitemOpen
  \bibfield  {author} {\bibinfo {author} {\bibfnamefont {I.}~\bibnamefont
  {Borukhov}}, \bibinfo {author} {\bibfnamefont {D.}~\bibnamefont {Andelman}},
  \ and\ \bibinfo {author} {\bibfnamefont {H.}~\bibnamefont {Orland}},\ }\href
  {\doibase 10.1007/s100510050513} {\bibfield  {journal} {\bibinfo  {journal}
  {Euro. Phys. J. B}\ }\textbf {\bibinfo {volume} {5}},\ \bibinfo {pages} {869}
  (\bibinfo {year} {1998})}\BibitemShut {NoStop}%
\bibitem [{\citenamefont {Wagner}\ \emph {et~al.}(2015)\citenamefont {Wagner},
  \citenamefont {Erdemci-Tandogan},\ and\ \citenamefont
  {Zandi}}]{adsorption2015}%
  \BibitemOpen
  \bibfield  {author} {\bibinfo {author} {\bibfnamefont {J.}~\bibnamefont
  {Wagner}}, \bibinfo {author} {\bibfnamefont {G.}~\bibnamefont
  {Erdemci-Tandogan}}, \ and\ \bibinfo {author} {\bibfnamefont
  {R.}~\bibnamefont {Zandi}},\ }\href {\doibase 10.1088/0953-8984/27/49/495101}
  {\bibfield  {journal} {\bibinfo  {journal} {J. Phys.:Condens. Matter}\
  }\textbf {\bibinfo {volume} {27}},\ \bibinfo {pages} {495101} (\bibinfo
  {year} {2015})}\BibitemShut {NoStop}%
\bibitem [{\citenamefont {Erdemci-Tandogan}\ \emph
  {et~al.}(2016{\natexlab{b}})\citenamefont {Erdemci-Tandogan}, \citenamefont
  {Wagner}, \citenamefont {van~der Schoot},\ and\ \citenamefont
  {Zandi}}]{Erdemci2016}%
  \BibitemOpen
  \bibfield  {author} {\bibinfo {author} {\bibfnamefont {G.}~\bibnamefont
  {Erdemci-Tandogan}}, \bibinfo {author} {\bibfnamefont {J.}~\bibnamefont
  {Wagner}}, \bibinfo {author} {\bibfnamefont {P.}~\bibnamefont {van~der
  Schoot}}, \ and\ \bibinfo {author} {\bibfnamefont {R.}~\bibnamefont
  {Zandi}},\ }\href {\doibase 10.1021/acs.jpcb.6b02712} {\bibfield  {journal}
  {\bibinfo  {journal} {J. Phys. Chem. B}\ }\textbf {\bibinfo {volume} {120}},\
  \bibinfo {pages} {6298} (\bibinfo {year} {2016}{\natexlab{b}})}\BibitemShut
  {NoStop}%
\bibitem [{\citenamefont {Janssen}\ \emph {et~al.}(2014)\citenamefont
  {Janssen}, \citenamefont {H\"artel},\ and\ \citenamefont {van
  Roij}}]{Janssen2014}%
  \BibitemOpen
  \bibfield  {author} {\bibinfo {author} {\bibfnamefont {M.}~\bibnamefont
  {Janssen}}, \bibinfo {author} {\bibfnamefont {A.}~\bibnamefont {H\"artel}}, \
  and\ \bibinfo {author} {\bibfnamefont {R.}~\bibnamefont {van Roij}},\ }\href
  {\doibase 10.1103/PhysRevLett.113.268501} {\bibfield  {journal} {\bibinfo
  {journal} {Phys. Rev. Lett.}\ }\textbf {\bibinfo {volume} {113}},\ \bibinfo
  {pages} {268501} (\bibinfo {year} {2014})}\BibitemShut {NoStop}%
\bibitem [{\citenamefont {de~Gennes}(1979)}]{deGennes1979}%
  \BibitemOpen
  \bibfield  {author} {\bibinfo {author} {\bibfnamefont {P.-G.}\ \bibnamefont
  {de~Gennes}},\ }\href@noop {} {\emph {\bibinfo {title} {Scaling concepts in
  polymer physics}}}\ (\bibinfo  {publisher} {Cornell University Press},\
  \bibinfo {year} {1979})\BibitemShut {NoStop}%
\bibitem [{\citenamefont {Siber}\ and\ \citenamefont
  {Podgornik}(2008)}]{Siber2008}%
  \BibitemOpen
  \bibfield  {author} {\bibinfo {author} {\bibfnamefont {A.}~\bibnamefont
  {Siber}}\ and\ \bibinfo {author} {\bibfnamefont {R.}~\bibnamefont
  {Podgornik}},\ }\href {\doibase 10.1103/PhysRevE.78.051915} {\bibfield
  {journal} {\bibinfo  {journal} {Phys. Rev. E}\ }\textbf {\bibinfo {volume}
  {78}},\ \bibinfo {pages} {051915} (\bibinfo {year} {2008})}\BibitemShut
  {NoStop}%
\bibitem [{\citenamefont {Shafir}\ \emph {et~al.}(2003)\citenamefont {Shafir},
  \citenamefont {Andelman},\ and\ \citenamefont {Netz}}]{Shafir}%
  \BibitemOpen
  \bibfield  {author} {\bibinfo {author} {\bibfnamefont {A.}~\bibnamefont
  {Shafir}}, \bibinfo {author} {\bibfnamefont {D.}~\bibnamefont {Andelman}}, \
  and\ \bibinfo {author} {\bibfnamefont {R.~R.}\ \bibnamefont {Netz}},\ }\href
  {\doibase 10.1063/1.1580798} {\bibfield  {journal} {\bibinfo  {journal} {J.
  Chem. Phys.}\ }\textbf {\bibinfo {volume} {119}},\ \bibinfo {pages} {2355}
  (\bibinfo {year} {2003})}\BibitemShut {NoStop}%
\bibitem [{\citenamefont {Lubensky}\ and\ \citenamefont
  {Isaacson}(1979)}]{Lubensky}%
  \BibitemOpen
  \bibfield  {author} {\bibinfo {author} {\bibfnamefont {T.~C.}\ \bibnamefont
  {Lubensky}}\ and\ \bibinfo {author} {\bibfnamefont {J.}~\bibnamefont
  {Isaacson}},\ }\href {\doibase 10.1103/PhysRevA.20.2130} {\bibfield
  {journal} {\bibinfo  {journal} {Phys. Rev. A}\ }\textbf {\bibinfo {volume}
  {20}},\ \bibinfo {pages} {2130} (\bibinfo {year} {1979})}\BibitemShut
  {NoStop}%
\bibitem [{\citenamefont {Nguyen}\ and\ \citenamefont
  {Bruinsma}(2006)}]{Nguyen-Bruinsma}%
  \BibitemOpen
  \bibfield  {author} {\bibinfo {author} {\bibfnamefont {T.~T.}\ \bibnamefont
  {Nguyen}}\ and\ \bibinfo {author} {\bibfnamefont {R.~F.}\ \bibnamefont
  {Bruinsma}},\ }\href {\doibase 10.1103/PhysRevLett.97.108102} {\bibfield
  {journal} {\bibinfo  {journal} {Phys. Rev. Lett.}\ }\textbf {\bibinfo
  {volume} {97}},\ \bibinfo {pages} {108102} (\bibinfo {year}
  {2006})}\BibitemShut {NoStop}%
\bibitem [{\citenamefont {Lee}\ and\ \citenamefont
  {Nguyen}(2008)}]{Lee-Nguyen}%
  \BibitemOpen
  \bibfield  {author} {\bibinfo {author} {\bibfnamefont {S.~I.}\ \bibnamefont
  {Lee}}\ and\ \bibinfo {author} {\bibfnamefont {T.~T.}\ \bibnamefont
  {Nguyen}},\ }\href {\doibase 10.1103/PhysRevLett.100.198102} {\bibfield
  {journal} {\bibinfo  {journal} {Phys. Rev. Lett.}\ }\textbf {\bibinfo
  {volume} {100}},\ \bibinfo {pages} {198102} (\bibinfo {year}
  {2008})}\BibitemShut {NoStop}%
\bibitem [{\citenamefont {Elleuch}\ \emph {et~al.}(1995)\citenamefont
  {Elleuch}, \citenamefont {Lequeux},\ and\ \citenamefont {Pfeuty}}]{Elleuch}%
  \BibitemOpen
  \bibfield  {author} {\bibinfo {author} {\bibfnamefont {K.}~\bibnamefont
  {Elleuch}}, \bibinfo {author} {\bibfnamefont {F.}~\bibnamefont {Lequeux}}, \
  and\ \bibinfo {author} {\bibfnamefont {P.}~\bibnamefont {Pfeuty}},\ }\href
  {\doibase 10.1051/jp1:1995140} {\bibfield  {journal} {\bibinfo  {journal} {J.
  Phys. I France}\ }\textbf {\bibinfo {volume} {5}},\ \bibinfo {pages} {465}
  (\bibinfo {year} {1995})}\BibitemShut {NoStop}%
\bibitem [{\citenamefont {de~Gennes}(1982)}]{deGennes}%
  \BibitemOpen
  \bibfield  {author} {\bibinfo {author} {\bibfnamefont {P.-G.}\ \bibnamefont
  {de~Gennes}},\ }\href@noop {} {\bibfield  {journal} {\bibinfo  {journal}
  {Macromolecules}\ }\textbf {\bibinfo {volume} {15}},\ \bibinfo {pages} {492}
  (\bibinfo {year} {1982})}\BibitemShut {NoStop}%
\bibitem [{\citenamefont {Ji}\ and\ \citenamefont {Hone}(1988)}]{Hone}%
  \BibitemOpen
  \bibfield  {author} {\bibinfo {author} {\bibfnamefont {H.}~\bibnamefont
  {Ji}}\ and\ \bibinfo {author} {\bibfnamefont {D.}~\bibnamefont {Hone}},\
  }\href {\doibase 10.1021/ma00186a049} {\bibfield  {journal} {\bibinfo
  {journal} {Macromolecules}\ }\textbf {\bibinfo {volume} {21}},\ \bibinfo
  {pages} {2600} (\bibinfo {year} {1988})}\BibitemShut {NoStop}%
\bibitem [{\citenamefont {Erdemci-Tandogan}\ \emph {et~al.}(2017)\citenamefont
  {Erdemci-Tandogan}, \citenamefont {Orland},\ and\ \citenamefont
  {Zandi}}]{GoncaPRL2017}%
  \BibitemOpen
  \bibfield  {author} {\bibinfo {author} {\bibfnamefont {G.}~\bibnamefont
  {Erdemci-Tandogan}}, \bibinfo {author} {\bibfnamefont {H.}~\bibnamefont
  {Orland}}, \ and\ \bibinfo {author} {\bibfnamefont {R.}~\bibnamefont
  {Zandi}},\ }\href {\doibase 10.1103/PhysRevLett.119.188102} {\bibfield
  {journal} {\bibinfo  {journal} {Phys. Rev. Lett.}\ }\textbf {\bibinfo
  {volume} {119}},\ \bibinfo {pages} {188102} (\bibinfo {year}
  {2017})}\BibitemShut {NoStop}%
\bibitem [{\citenamefont {Zandi}\ and\ \citenamefont {van~der
  Schoot}(2009)}]{Vanderschoot2009}%
  \BibitemOpen
  \bibfield  {author} {\bibinfo {author} {\bibfnamefont {R.}~\bibnamefont
  {Zandi}}\ and\ \bibinfo {author} {\bibfnamefont {P.}~\bibnamefont {van~der
  Schoot}},\ }\href {\doibase 10.1529/biophysj.108.137489} {\bibfield
  {journal} {\bibinfo  {journal} {Biophys. J.}\ }\textbf {\bibinfo {volume}
  {96}},\ \bibinfo {pages} {9} (\bibinfo {year} {2009})}\BibitemShut {NoStop}%
\bibitem [{\citenamefont {Siber}\ \emph {et~al.}(2010)\citenamefont {Siber},
  \citenamefont {Zandi},\ and\ \citenamefont {Podgornik}}]{SiberZandi2010}%
  \BibitemOpen
  \bibfield  {author} {\bibinfo {author} {\bibfnamefont {A.}~\bibnamefont
  {Siber}}, \bibinfo {author} {\bibfnamefont {R.}~\bibnamefont {Zandi}}, \ and\
  \bibinfo {author} {\bibfnamefont {R.}~\bibnamefont {Podgornik}},\ }\href
  {\doibase 10.1103/PhysRevE.81.051919} {\bibfield  {journal} {\bibinfo
  {journal} {Phys. Rev. E}\ }\textbf {\bibinfo {volume} {81}},\ \bibinfo
  {pages} {051919} (\bibinfo {year} {2010})}\BibitemShut {NoStop}%
\bibitem [{\citenamefont {Borodavka}\ \emph {et~al.}(2016)\citenamefont
  {Borodavka}, \citenamefont {Singaram}, \citenamefont {Stockley},
  \citenamefont {Gelbart}, \citenamefont {Ben-Shaul},\ and\ \citenamefont
  {Tuma}}]{BORODAVKA2016}%
  \BibitemOpen
  \bibfield  {author} {\bibinfo {author} {\bibfnamefont {A.}~\bibnamefont
  {Borodavka}}, \bibinfo {author} {\bibfnamefont {S.}~\bibnamefont {Singaram}},
  \bibinfo {author} {\bibfnamefont {P.}~\bibnamefont {Stockley}}, \bibinfo
  {author} {\bibfnamefont {W.}~\bibnamefont {Gelbart}}, \bibinfo {author}
  {\bibfnamefont {A.}~\bibnamefont {Ben-Shaul}}, \ and\ \bibinfo {author}
  {\bibfnamefont {R.}~\bibnamefont {Tuma}},\ }\href {\doibase
  https://doi.org/10.1016/j.bpj.2016.10.014} {\bibfield  {journal} {\bibinfo
  {journal} {Biophysical Journal}\ }\textbf {\bibinfo {volume} {111}},\
  \bibinfo {pages} {2077 } (\bibinfo {year} {2016})}\BibitemShut {NoStop}%
\end{thebibliography}%
\end{document}